\documentclass[a4paper,12pt]{article}
\usepackage{lineno,hyperref}
\usepackage[english]{babel}
\usepackage{graphicx}
\usepackage{authblk}
\usepackage{subfigure} 
\usepackage{mathtools}                
\usepackage{float}
\usepackage{numcompress}
\begin{document}

\title{ Bin Size Independence in Intra-day Seasonalities for Relative Prices }

\author[1,2]{Esteban Guevara Hidalgo}
\affil[1]{\small Institut Jacques Monod, CNRS UMR 7592, Universit\'e Paris Diderot, Sorbonne Paris Cit\'e, F-750205, Paris, France}
\affil[2]{\small Laboratoire de Probabilit\'es et Mod\`eles Al\'eatoires, Sorbonne Paris Cit\'e, UMR 7599 CNRS, Universit\'e Paris Diderot, 75013 Paris, France \texttt{esteban\_guevarah@hotmail.com}}

\date{}

\maketitle

\begin{abstract}
In this paper we perform a statistical analysis over the returns and relative prices of the CAC $40$ and the S\&P $500$ with the purpose of analyzing the intra-day seasonalities of single and cross-sectional stock dynamics. In order to do that, we characterized the dynamics of a stock (or a set of stocks) by the evolution of the moments of its returns (and relative prices) during a typical day. We show that these intra-day seasonalities are independent of the size of the bin, and the index we consider, (but characteristic for each index) for the case of the relative prices but not for the case of the returns.  Finally, we suggest how this bin size independence could be used to characterize ``atypical days'' for indexes and ``anomalous behaviours'' in stocks. 
\end{abstract}


\section{Introduction}
From the statistical study of financial time series have arisen a set of  properties or empirical laws sometimes called ``stylized facts'' or seasonalities. These properties have the characteristic of being  common and persistent across different markets, time periods and assets \cite{1, 2, 3, 4, 5, 6, 7}.
As it has been suggested \cite{7}, the reason why these ``patterns''  appear could be because markets operate in synchronization with human activities which leave a trace in the financial time series.
However using the ``right clock'' might be of primary importance when dealing with statistical properties and the patterns could vary depending if we use daily data or intra-day data and event time, trade time or arbitrary intervals of time (e.g. $T = 1$, $5$, $15$ minutes, etc.). 
For example, it is a well-known fact that empirical distributions of  financial returns and log-returns are fat tailed \cite{8, 9}, however as one increases the time scale the fat-tail property becomes less pronounced  and the distribution approach the Gaussian form \cite{10}. As was stated in \cite{4}, the fact that the shape of the distribution changes with time makes it clear that the random process underlying prices must have a non-trivial temporal structure.
In a previous work Allez et al. \cite{7} established several new stylized facts concerning the intra-day seasonalities of single and cross-sectional stock dynamics. This dynamics is characterized by the evolution of the moments of its returns  during a typical day.  Following the same approach, we show the bin size dependence of these patterns for the case of returns and, motivated by the work of Kaisoji \cite{11}, we extend the analysis to relative prices and show how in this case, these patterns are independent of the size of the bin, also independent of the index we consider but characteristic for each index. These facts could be used in order to detect an anomalous behaviour during the day, like market crashes or intra-day bubbles \cite{11, 12}. The present work is completely empirical but it could offer signs of the  underlying stochastic process that governs the financial time series. 

\section{Definitions}
The data consists in two sets of intra-day high frequency time series, the CAC $40$ and the S\&P $500$. For each of the $D = 22$ days of our period of analysis (March $2011$), we dispose with the evolution of the prices of each of the stocks that composes our indexes during a specific day from $10:00$ a.m. to $16:00$ p.m.  The
main reasons why we chose to work with these two indexes are: The number of stocks that
compose them ($N_{1}=40$ and $N_{2}=500$), the time gap between their respective markets and the
different range of stock prices (between $5$ and $600$ USD for the S\&P $500$
and between $5$ and $145$ EU for the CAC $40$). 

\begin{figure}[h!]
\centering
\includegraphics [scale=0.5] {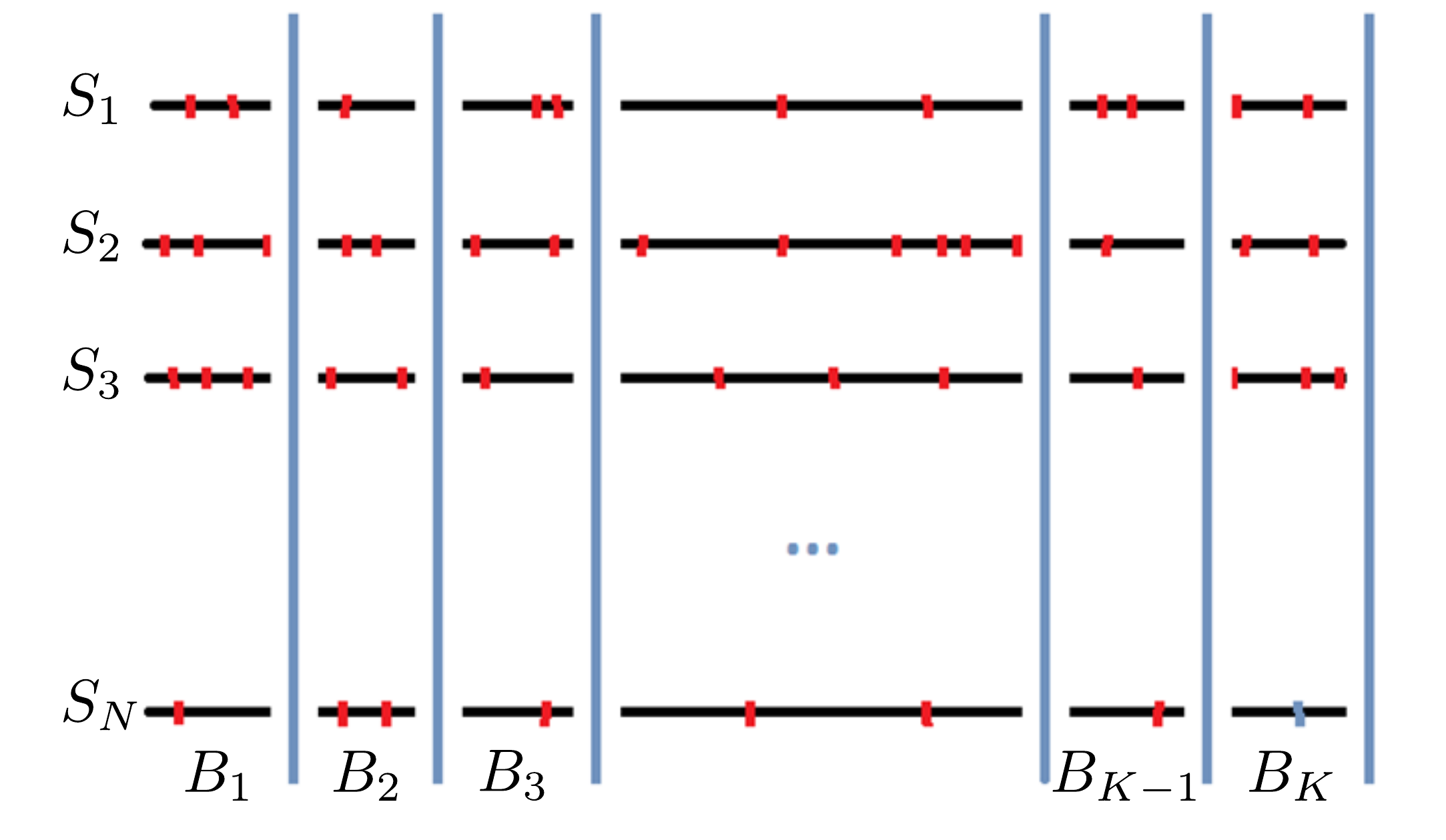}
\caption{Intra-day asynchronous financial time series. $S_{i}$ are the stocks and $B_{k}$ are bins. The asynchronous prices are show in red and the bin limit in blue.}
\label{fig:Fig1}
\end{figure}

As the changes in prices are not synchronous between different stocks (figure~\ref{fig:Fig1}), we manipulated our original data in order to construct a new homogeneous matrix $P_{D}^{(j)}$ of ``bin prices''. In order to do this, we divided our daily time interval $[10:00,16:00]$ in $K$ bins of size $T$ (minutes), i.e. $B_{1} = [10:00,10:00 + T]$, $B_{2} = [10:00 + T,10:00 + 2T],..., B_{K} = [16:00 - T,16:00]$, where the right endpoints of these intervals are called ``bin limits''. For a particular day $j$, the prices that conform the matrix $P_{D}^{(j)}$ are given by the last prices that stock $i$ reaches just before a specific bin limit.

Each row in the matrix below represents the evolution of the prices of a
particular stock as function of the bins. For example, the element $%
(P_{D})_{ik}^{(j)}$, represents the price for a particular day $j$ of the stock $i$ and just before the bin limit of the bin $B_{k}$.
\begin{equation}
P_{D}^{(j)}=%
\begin{pmatrix}
P_{11}^{(j)} & P_{12}^{(j)} & ... & ... & ... & P_{1K}^{(j)} \\ 
P_{21}^{(j)} & P_{22}^{(j)} & ... & ... & ... & P_{2K}^{(j)} \\ 
... & ... & ... & ... & ... & ... \\ 
... & ... & ... & P_{ik}^{(j)} & ... & ... \\ 
... & ... & ... & ... & ... & ... \\ 
P_{N1}^{(j)} & P_{N2}^{(j)} & ... & ... & .. & P_{NK}^{(j)}%
\end{pmatrix}
\label{A1}
\end{equation}%

In a similar way, we can construct the matrix $P_{S}^{(i)}$ for each of the $%
i=1,...,N_{1,2}$ stocks. $(P_{S})_{jk}^{(i)}$ is the price of the stock $(i)$ in the day $j$ and just before the bin limit of the bin $B_{k}$.%
\begin{equation}
P_{S}^{(i)}=%
\begin{pmatrix}
P_{11}^{(i)} & P_{12}^{(i)} & ... & ... & ... & P_{1K}^{(i)} \\ 
P_{21}^{(i)} & P_{22}^{(i)} & ... & ... & ... & P_{2K}^{(i)} \\ 
... & ... & ... & ... & ... & ... \\ 
... & ... & ... & P_{jk}^{(i)} & ... & ... \\ 
... & ... & ... & ... & ... & ... \\ 
P_{D1}^{(i)} & P_{D2}^{(i)} & ... & ... & .. & P_{DK}^{(i)}%
\end{pmatrix}
\label{A2}
\end{equation}

\bigskip 

In the following and for simplicity, we will refer to the price $P$ of a particular stock $i = \alpha $ during a particular day $j = t$ and just before the bin limit of the bin $B_{k}$ as $P_{\alpha }(k,t)$ where $P_{\alpha }(k,t) = P_{tk}^{(\alpha)}=P_{\alpha k}^{(t)}$. 

We will perform our statistical analysis over the variable $x_{\alpha }(k,t)$ that can be computed from the matrices above. For our interests we will be working with returns 

\begin{equation}
x_{\alpha }^{(1)}(k,t)=\frac{P_{\alpha }(k+1,t)-P_{\alpha }(k,t)}{P_{\alpha
}(k,t)}  \label{1}
\end{equation}%
and relative prices \cite{11, 12}%
\begin{equation}
x_{\alpha }^{(2)}(k,t)=\frac{P_{\alpha }(k,t) - P_{\alpha }(1,t)}{P_{\alpha }(1,t)}
\label{2}
\end{equation}

The single or collective stock dynamics is characterized by the evolution of the moments of the returns (or relative prices). Below, we show how we computed these
moments \cite{7}.  

\subsection{Single Stock Properties}

The distribution of the stock\ $\alpha $\ in bin $k$ is characterized by its
four first moments: mean $\mu _{\alpha }(k)$, standard deviation
(volatility) $\sigma _{\alpha }(k),$ skewness $\zeta _{\alpha }(k)$ and
kurtosis $\kappa _{\alpha }(k)$ defined as%
\begin{eqnarray}
\mu _{\alpha }(k) &=&\left\langle x_{\alpha }(k,t)\right\rangle   \label{3}
\\
\sigma _{\alpha }^{2}(k) &=&\left\langle x_{\alpha }^{2}(k,t)\right\rangle
-\mu _{\alpha }^{2}(k)  \label{4} \\
\zeta _{\alpha }(k) &=&\frac{6}{\sigma _{\alpha }(k)}(\mu _{\alpha
}(k)-m_{\alpha }(k))  \label{5} \\
\kappa _{\alpha }(k) &=&24\left( 1-\sqrt{\frac{\pi }{2}}\frac{\left\langle
|x_{\alpha }(k,t)-\mu _{\alpha }(k)|\right\rangle }{\sigma _{\alpha }(k)}%
\right) +\zeta _{\alpha }^{2}(k)  \label{6}
\end{eqnarray}%
where $m_{\alpha }(k)$ is the median of all values of $x_{\alpha }(k,t)$ and time averages for a given stock in a given bin are expressed with angled brackets $\langle...\rangle$.

\subsection{Cross-Sectional Stock Properties}
The cross-sectional distributions (i.e. the dispersion of the values of the
variable $x$ of the $N$ stocks for a given bin $k$ in a given day $t$) are
also characterized by the four first moments%
\begin{eqnarray}
\mu _{d}(k,t) &=&\left[ x_{\alpha }(k,t)\right]   \label{7} \\
\sigma _{d}^{2}(k,t) &=&\left[ x_{\alpha }^{2}(k,t)\right] -\mu _{d}^{2}(k,t)
\label{8} \\
\zeta _{d}(k,t) &=&\frac{6}{\sigma _{d}(k,t)}(\mu _{d}(k,t)-m_{d}(k,t))
\label{9} \\
\kappa _{d}(k) &=&24\left( 1-\sqrt{\frac{\pi }{2}}\frac{\left[ |x_{\alpha
}(k,t)-\mu _{\alpha }(k)|\right] }{\sigma _{d}(k)}\right)   \label{10}
\end{eqnarray}%
where $m_{d}(k,t)$ is the median of all the $N$ values of the variable $x$
for a given $(k,t)$ and the square brackets $\left[...\right]$ represent averages over the ensemble
of stocks in a given bin and day. If $x_{\alpha }(k,t)$ are the returns, $\mu _{d}(k,t)$ can be seen as the return of an index
equi-weighted on all stocks.

\section{Intra-day Seasonalities for Returns}
The following results are in complete agreement with the results previously reported by \cite{5, 6, 7}.
\subsection{Single Stock Intra-day Seasonalities}
\begin{figure}[h!]
        \begin{center}
        \subfigure[MEAN]{\includegraphics [scale=0.4] {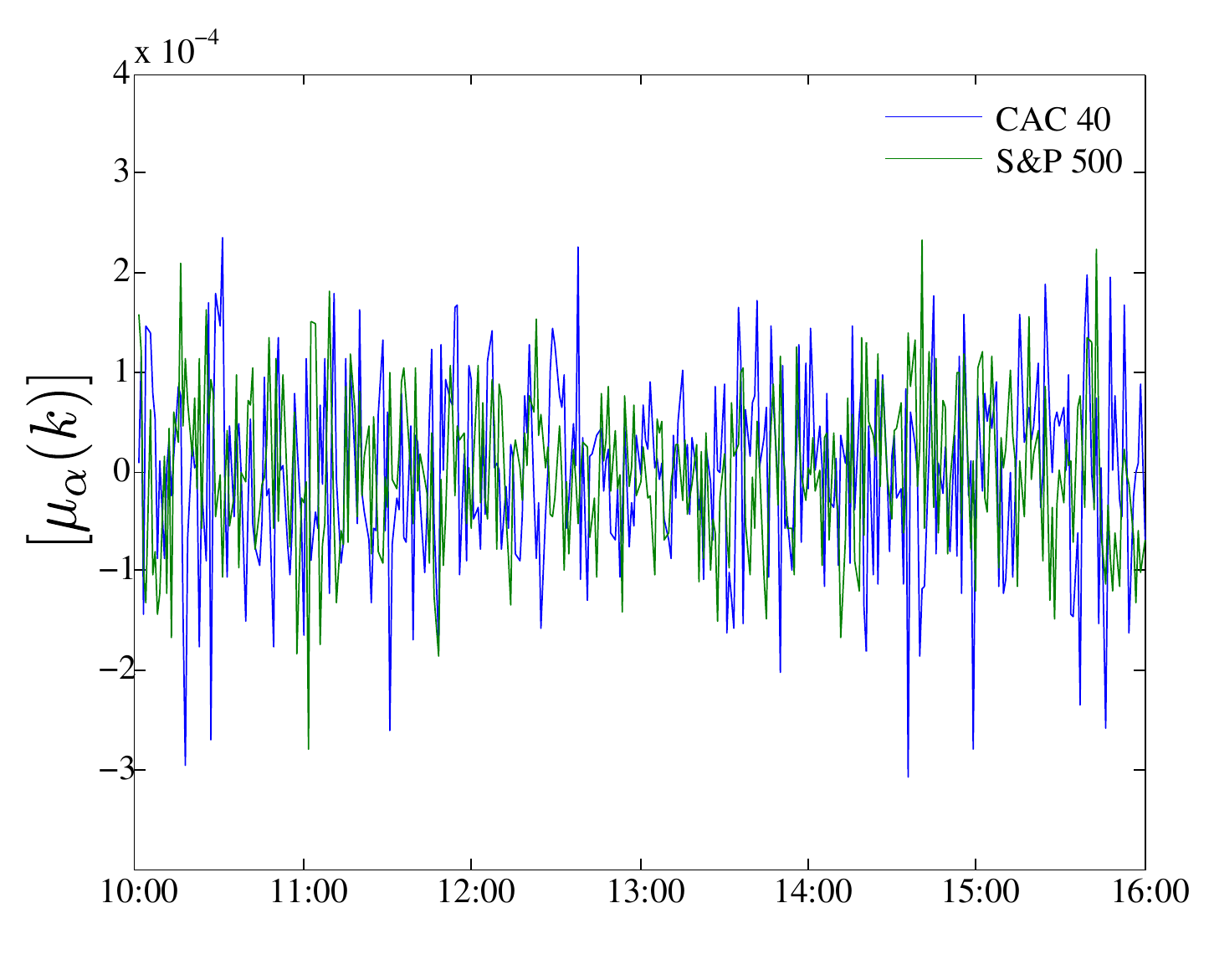}}
		\subfigure[VOLATILITY]{\includegraphics [scale=0.4] {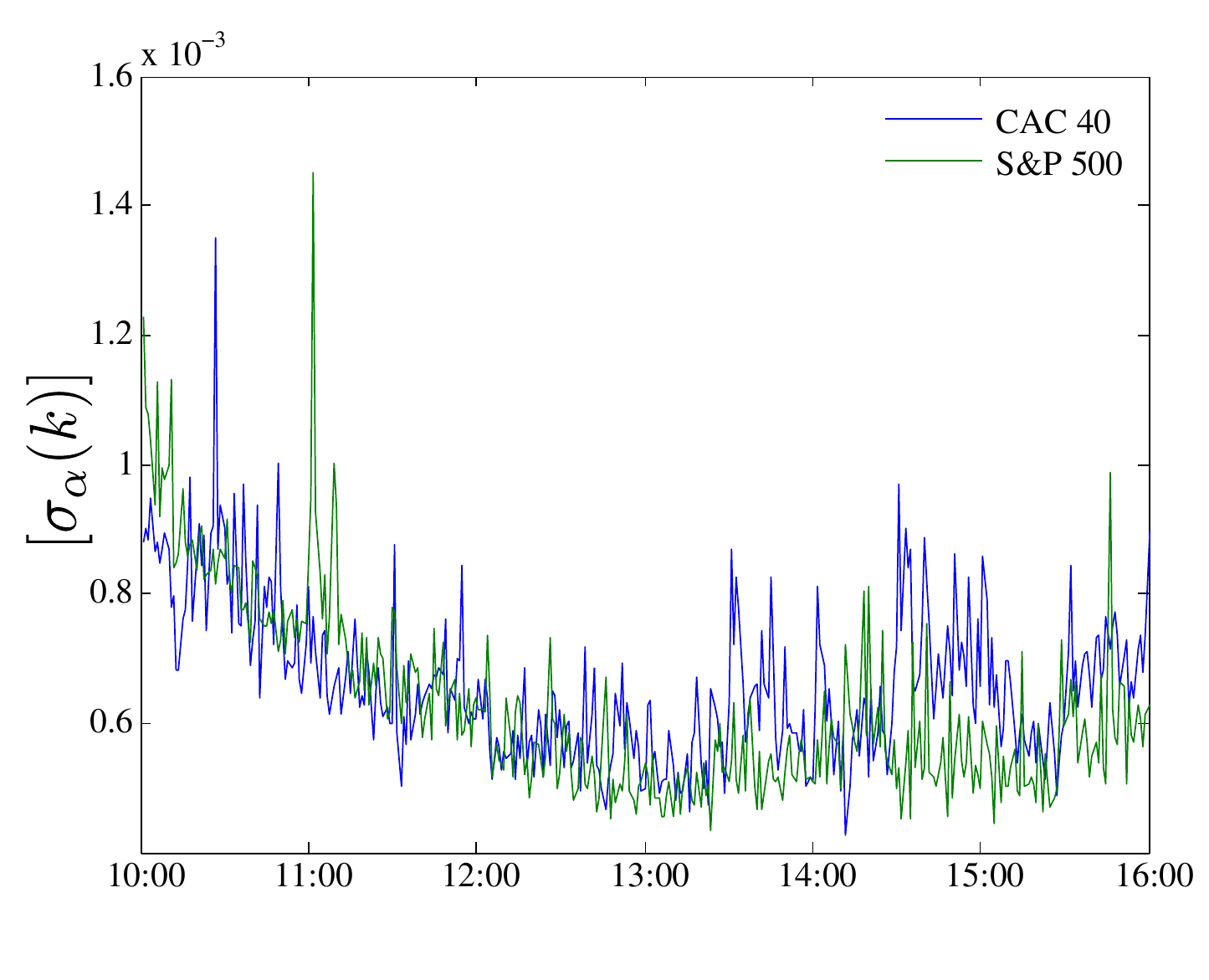}}\\ 
        \subfigure[SKEWNESS]{\includegraphics [scale=0.4] {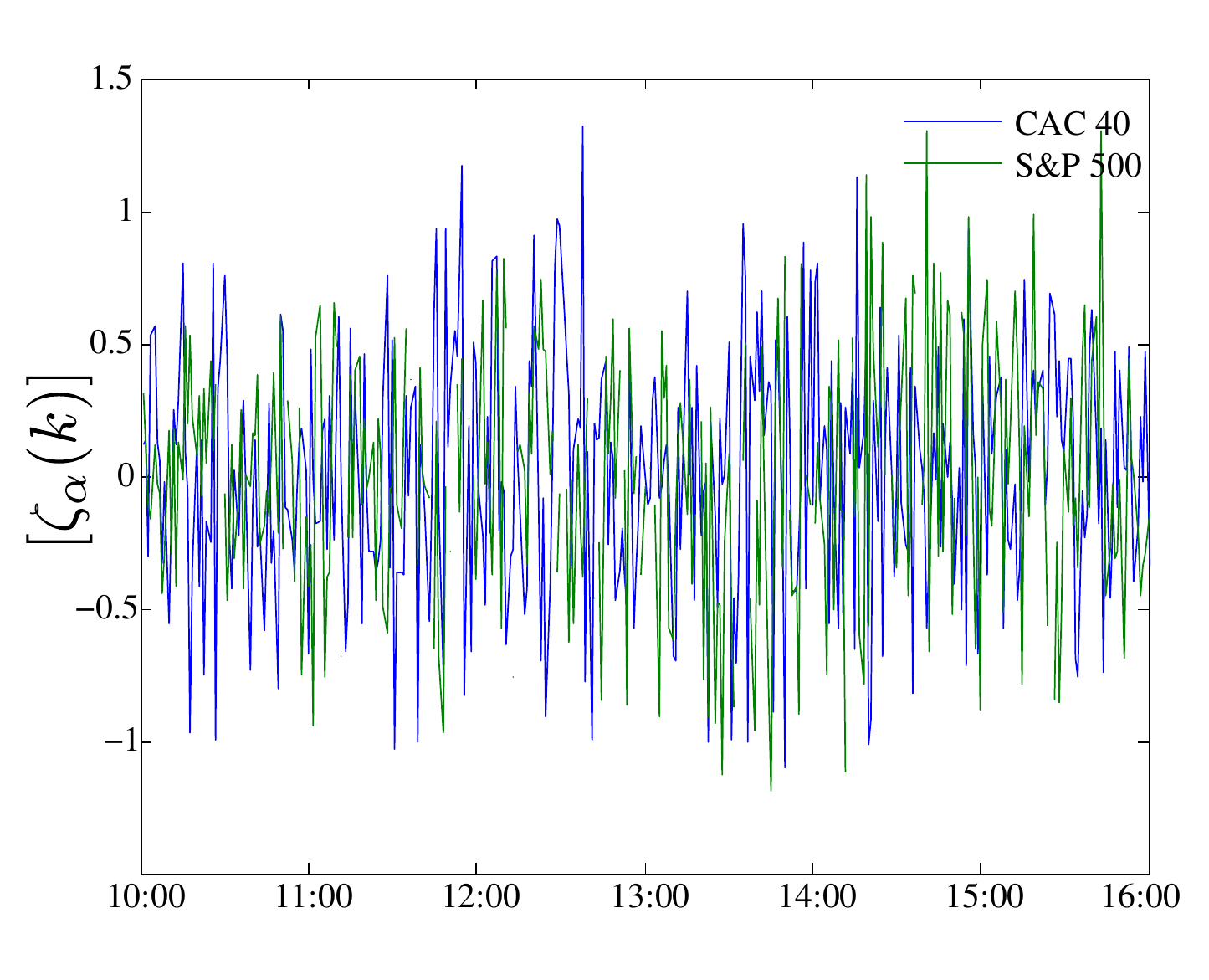}}
        \subfigure[KURTOSIS]{\includegraphics [scale=0.4] {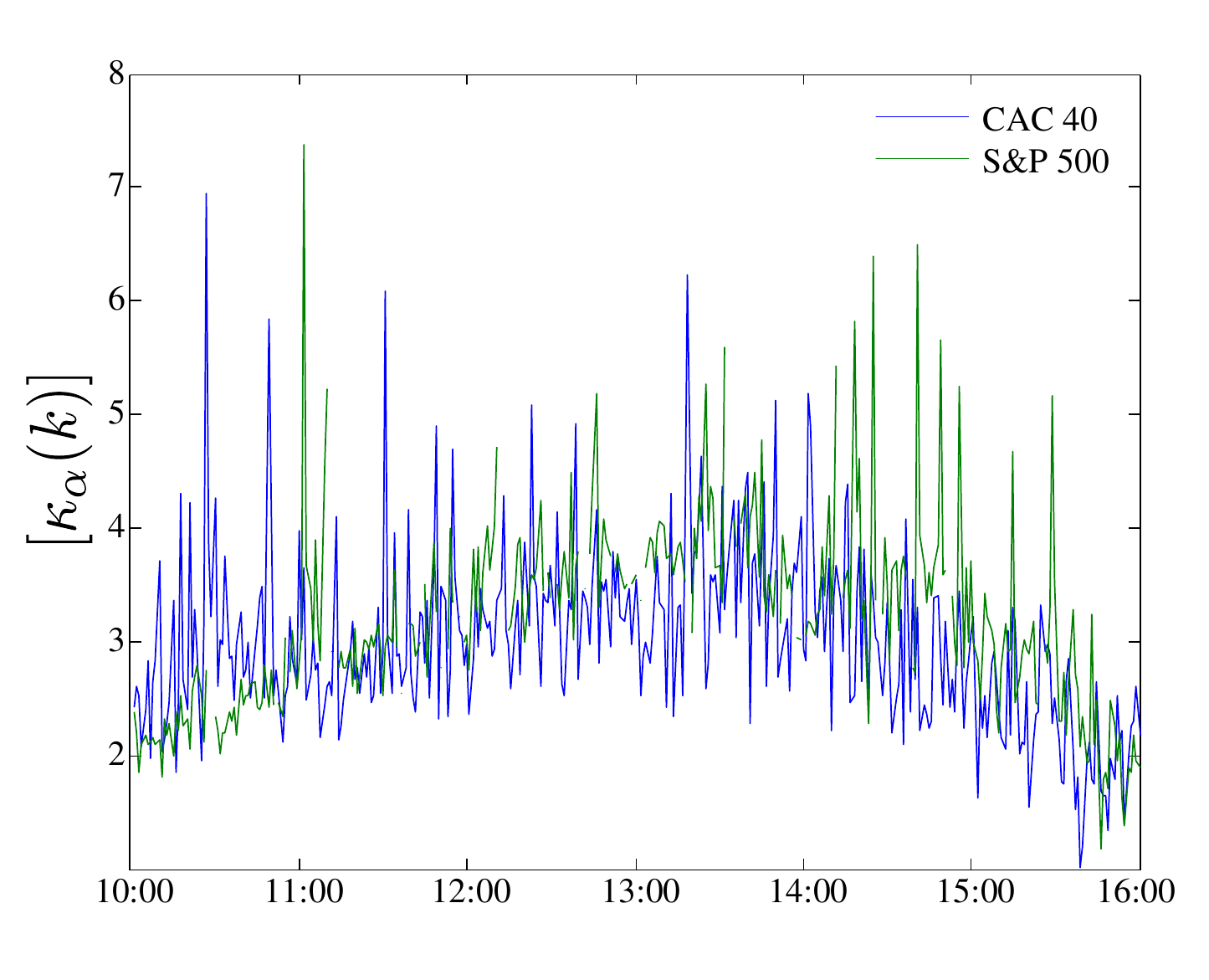}}                  
    	\end{center}
    \caption{Single Stock Intra-day Seasonalities: Stock average of the single stock mean, volatility, skewness and kurtosis for the CAC\ $40$ (blue) and the S\&P $500$ (green). $T=1$ minute bin.}
\label{fig:Fig2}
\end{figure}

Figure~\ref{fig:Fig2} shows the stock average of the single stock mean $\left[\mu_{\alpha }(k)\right]$, volatility $\left[\sigma _{\alpha }(k)\right]$, skewness $\left[\zeta_{\alpha }(k)\right]$ and kurtosis $\left[\kappa_{\alpha }(k)\right]$ for the CAC\ $40$ (blue) and the S\&P $500$ (green), and $T=1$ minute bin. As can be seen in figure~\ref{fig:Fig2}(a), the mean tends to be small (in the order of $10^{-4}$) and noisy around zero. The average volatility reveals the well known U-shaped pattern (figure~\ref{fig:Fig2}(b)), high at the opening of the day, decreases during the day and increases again at the end of the day. The average skewness (figure~\ref{fig:Fig2}(c)) is also noisy around zero. The average kurtosis exhibits an inverted U-pattern (figure~\ref{fig:Fig2}(d)), it increases from around  $2$ at the beginning of the day to around $4$ at mid day, and decreases again during the rest of the day.

\subsection{Cross-Sectional Intra-day Seasonalities}
\begin{figure}[h!]
        \begin{center}
        \subfigure[MEAN]{\includegraphics [scale=0.4] {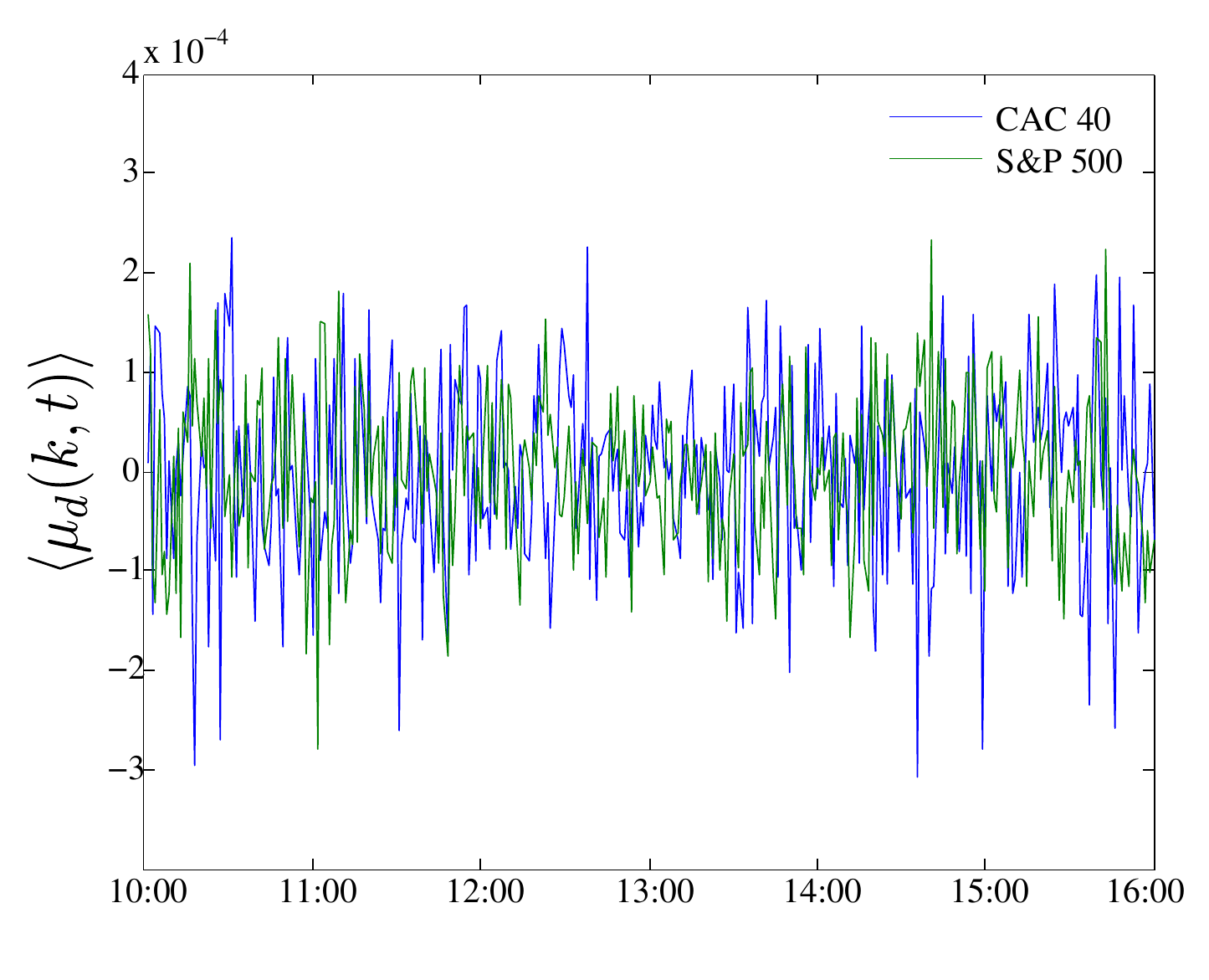}}
		\subfigure[VOLATILITY]{\includegraphics [scale=0.4] {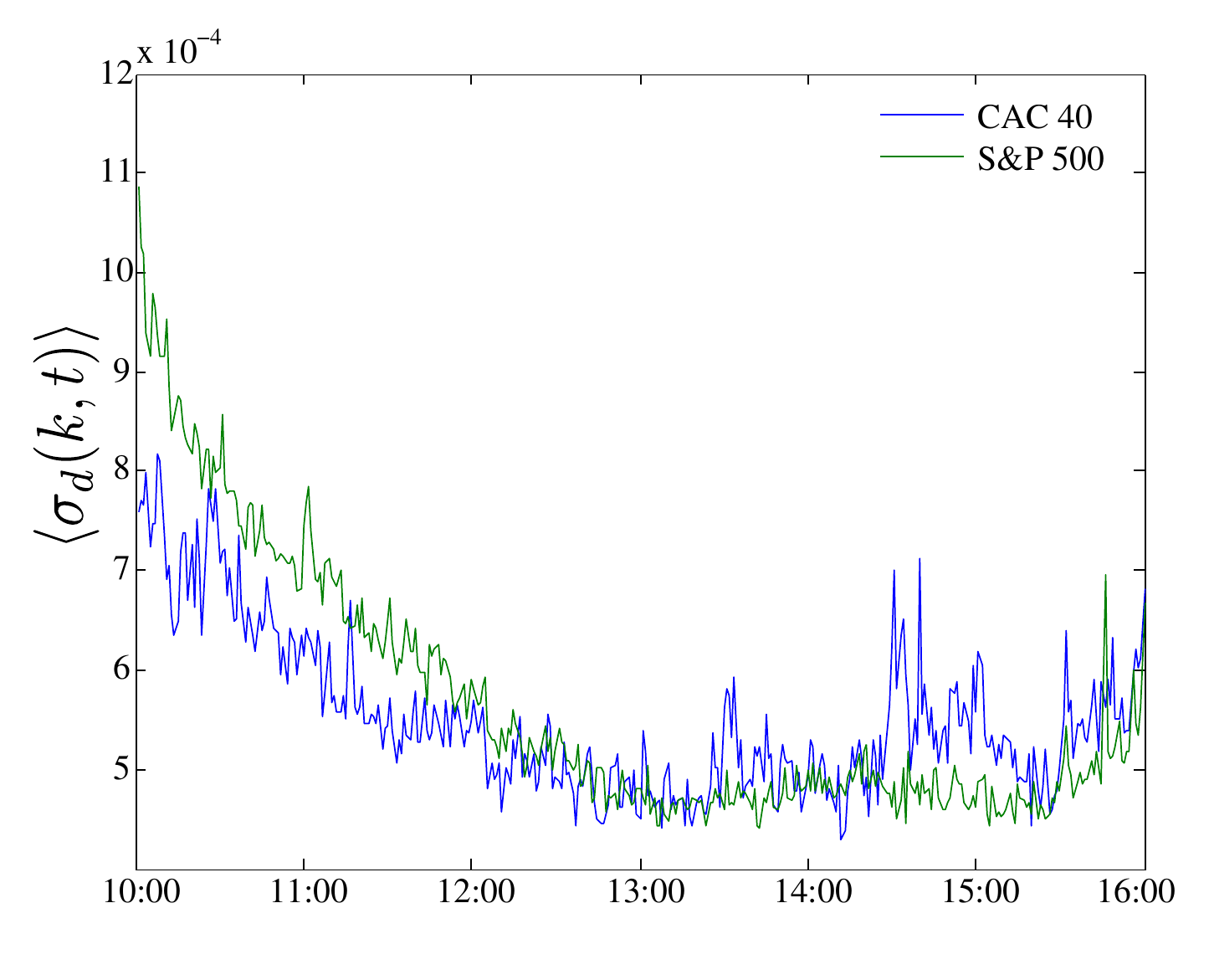}}\\ 
        \subfigure[SKEWNESS]{\includegraphics [scale=0.4] {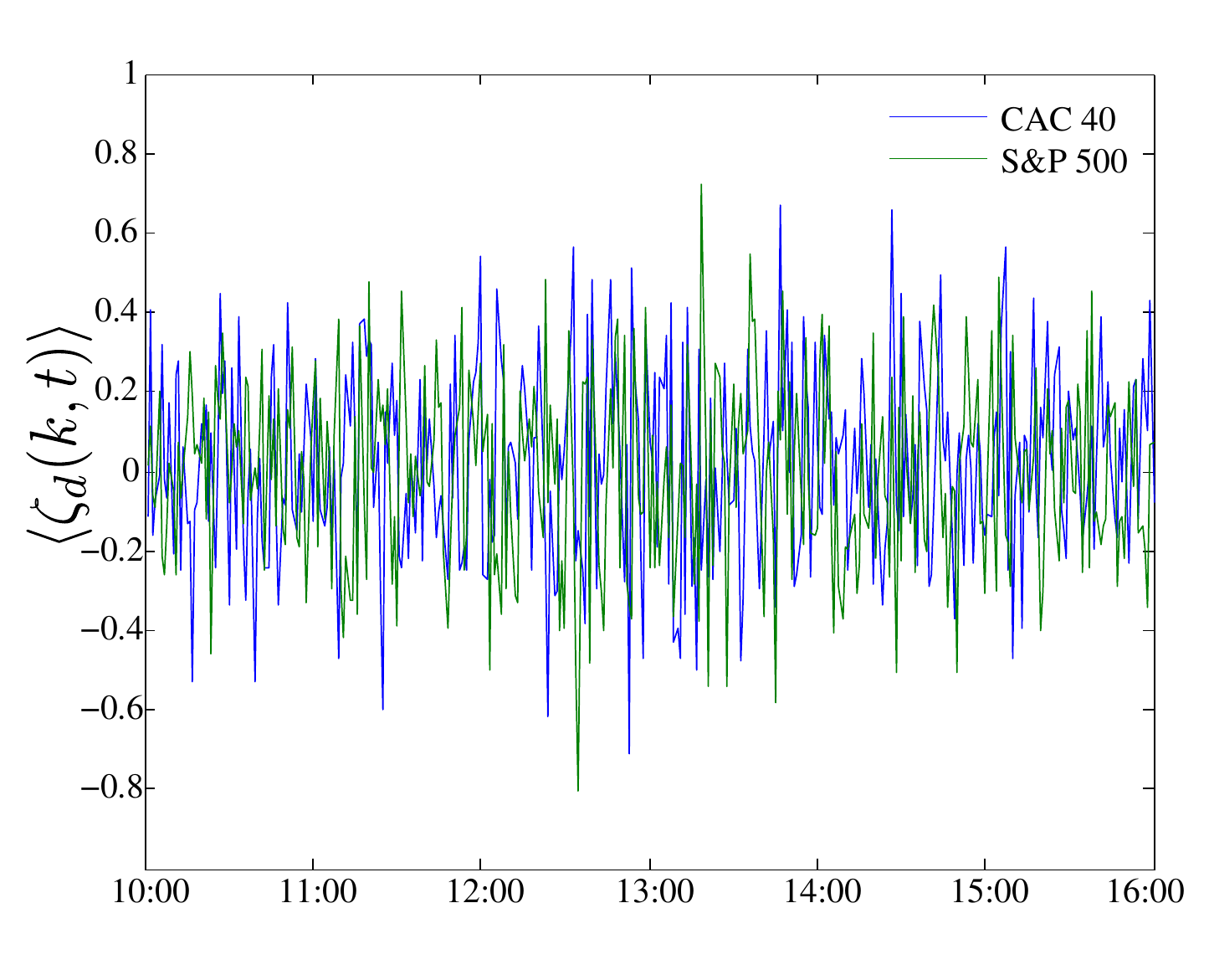}}
        \subfigure[KURTOSIS]{\includegraphics [scale=0.4] {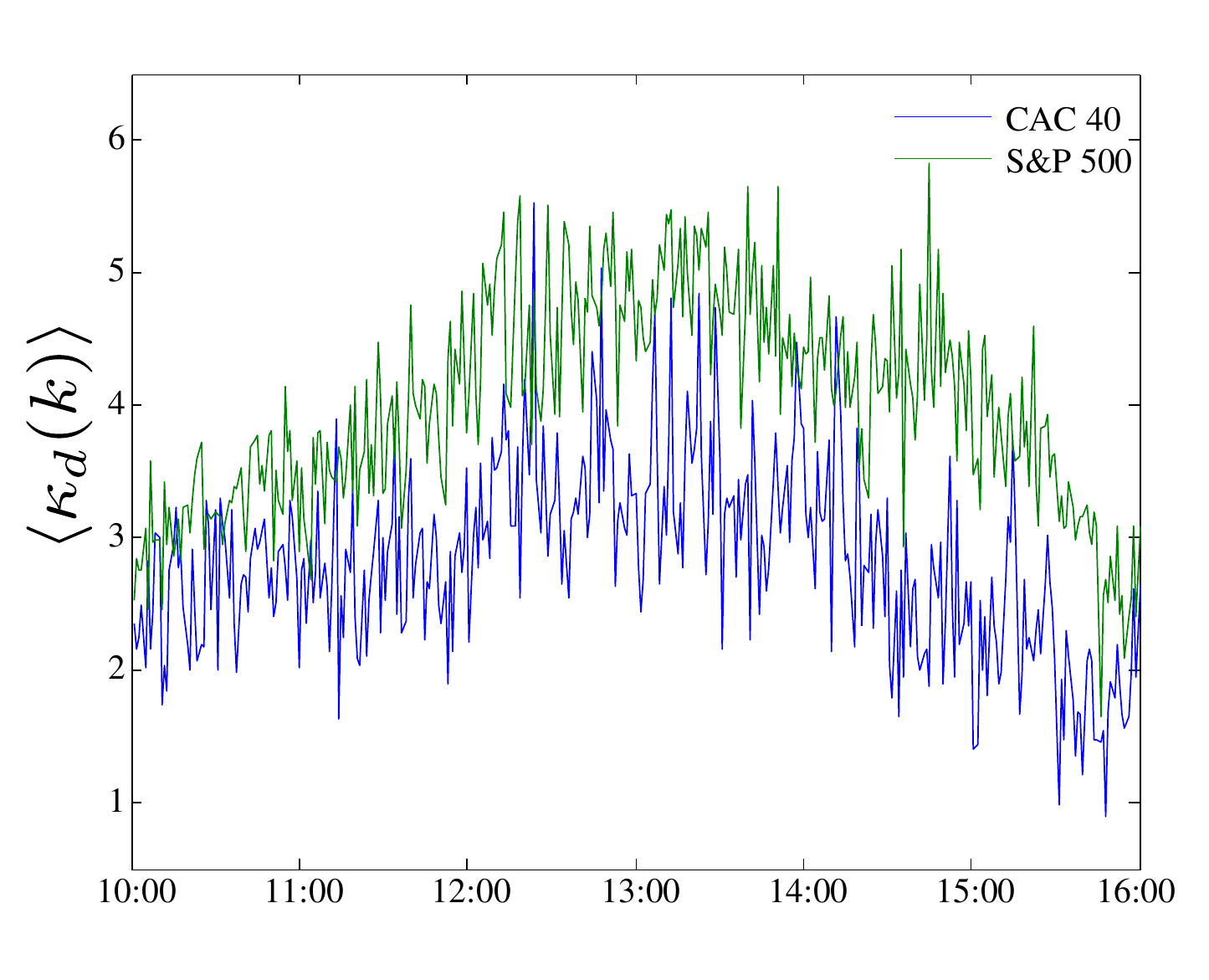}}                  
    	\end{center}
    \caption{Cross-Sectional Intra-day Seasonalities: Time average of the cross-sectional mean, volatility, skewness and kurtosis for the CAC\ $40$ (blue) and the S\&P $500$ (green), and $T=1$ minute bin.}
\label{fig:Fig3}
\end{figure}

As the time average of the cross sectional mean is equal to the stock average of the single stock mean, the result we show in figure~\ref{fig:Fig3}(a) is exactly the same as the one shown in figure~\ref{fig:Fig2}(a). The time average of the cross sectional volatility $\langle\sigma _{d}(k,t)\rangle$ (figure~\ref{fig:Fig3}(b)) reveals a U-shaped pattern very similar to the stock average volatility, but less noisy (less pronounced peaks). The dispersion of stocks is stronger at the beginning of the day and decreases as the day proceeds. The average skewness $\langle\zeta _{d}(k,t)\rangle$ is noisy around zero without any particular pattern (figure~\ref{fig:Fig3}(c)). The cross sectional kurtosis $\langle\kappa _{d}(k)\rangle$ (figure~\ref{fig:Fig3}(d)) also exhibits an inverted U-pattern as in the case of the single stock kurtosis. It increases from around $2.5$ at the beginning of the day to around $4.5$ at mid day, and decreases again during the rest of the day. This means that at the beginning of the day the cross-sectional distribution of returns is on average closer to Gaussian.%

\subsection{U-Pattern Volatilities}
\begin{figure}[h!]
        \begin{center}
        \subfigure[$T=1$ minute bin]{\includegraphics [scale=0.4] {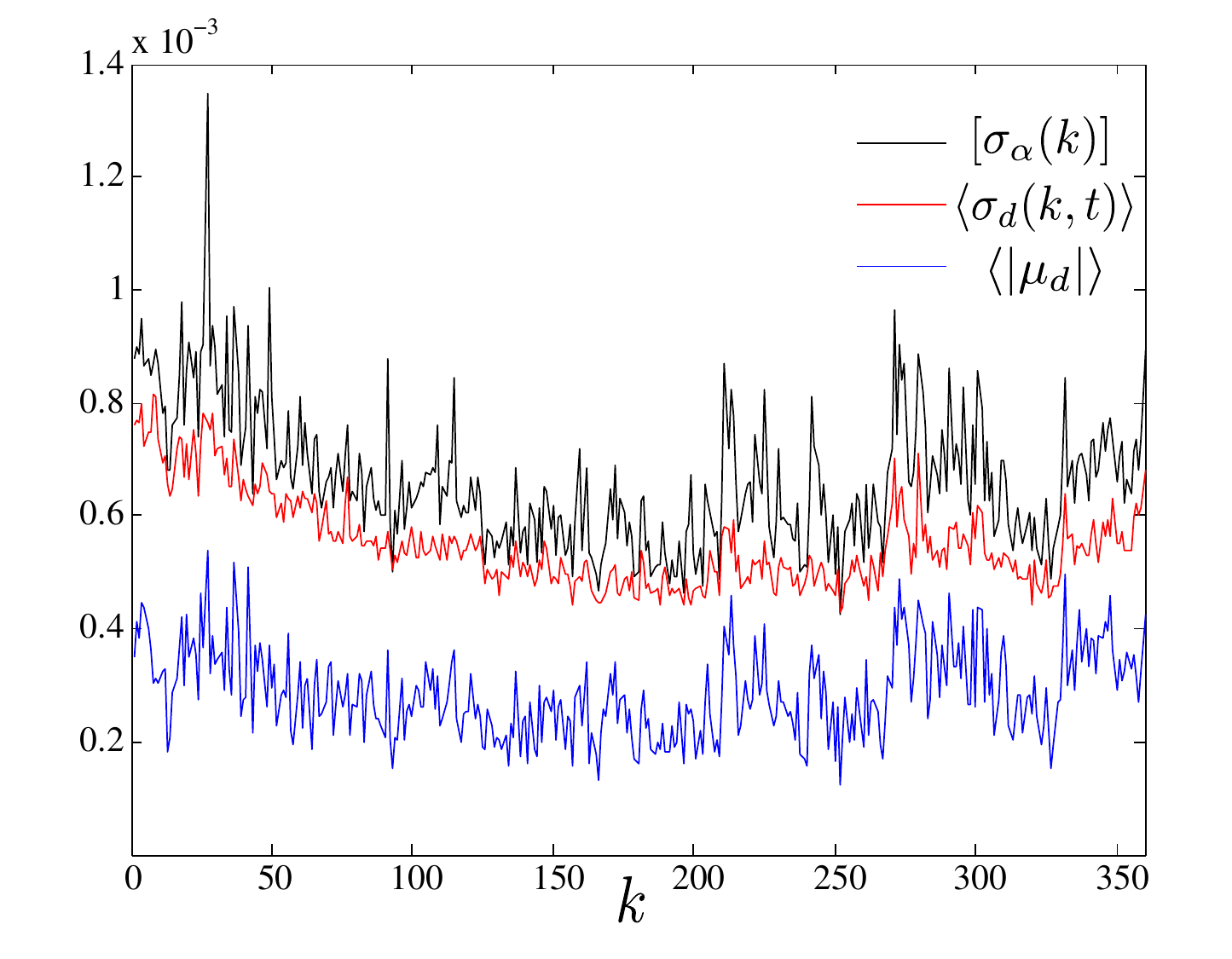}}
		\subfigure[$T=5$ minute bin]{\includegraphics [scale=0.4] {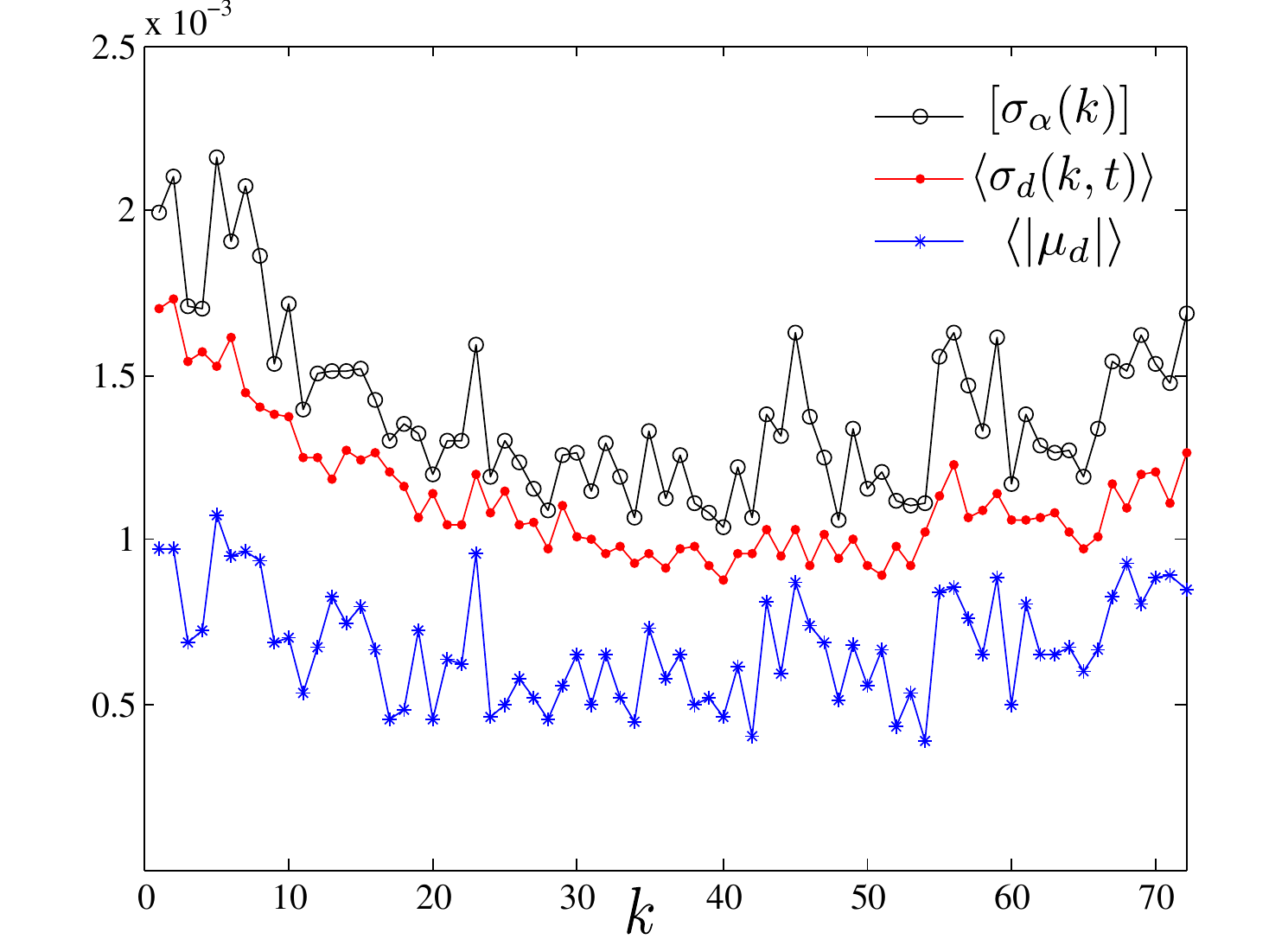}}           
    	\end{center}
    \caption{U-Pattern Volatilities: Stock average of single stock volatility $\left[\sigma _{\alpha }(k)\right]$ (black), time average of the cross sectional volatility $\langle\sigma _{d}(k,t)\rangle$ (red) and the average absolute value of the equi-weighted index return $\langle|\mu _{d}|\rangle$ (blue) for the CAC\ $40$. Similar results were obtained for the S\&P $500$.}
\label{fig:Fig4}
\end{figure}

In figure~\ref{fig:Fig4}, we compare the stock average of single stock volatility $\left[\sigma _{\alpha }(k)\right]$ (black), the time average of the cross-sectional volatility $ \langle \sigma _{d}(k,t)\rangle$ (red) and the average absolute value of the equi-weighted index return $\langle|\mu _{d}|\rangle$ (blue) for the CAC\ $40$, and for $T=1$ (left) and $T=5$ minute bin (right). Similar results were obtained for the S\&P $500$. As can be seen, the average absolute value of the equi-weighted index return also exhibits a U-shaped pattern and it is a proxy for the index volatility. One thing that results interesting to observe is that the values of these volatilities actually depends of the size of the bin that we consider. For $T=5$ minute bin, the volatilities double the values found for $T=1$ minute bin (we will discuss this result in the next sections).

\subsection{Intra-day Seasonalities in the Stock Correlation}
In order to compute the correlation between stocks, we first normalize the returns by the dispersion of the corresponding bin \cite{7} i.e.,
\begin{equation}
\widehat{x}_{\alpha}(k,t) = x_{\alpha }^{(1)}(k,t)/\sigma _{d}(k,t)
\label{11}
\end{equation}%
The $N\times N$ correlation matrix for a given bin $k$ would be given by%
\begin{equation}
C_{\alpha \beta }(k)=\frac{\left\langle \widehat{x}_{\alpha }(k,t)%
\widehat{x}_{\beta}(k,t)\right\rangle -\left\langle \widehat{x}_{\alpha}(k,t)\right\rangle \left\langle \widehat{x}_{\beta}(k,t)\right\rangle }{\sigma _{\alpha}(k)\sigma _{\beta}(k)} \label{12}
\end{equation}

\begin{figure}[h!]
        \begin{center}
        \subfigure[]
        {\includegraphics [scale=0.4] {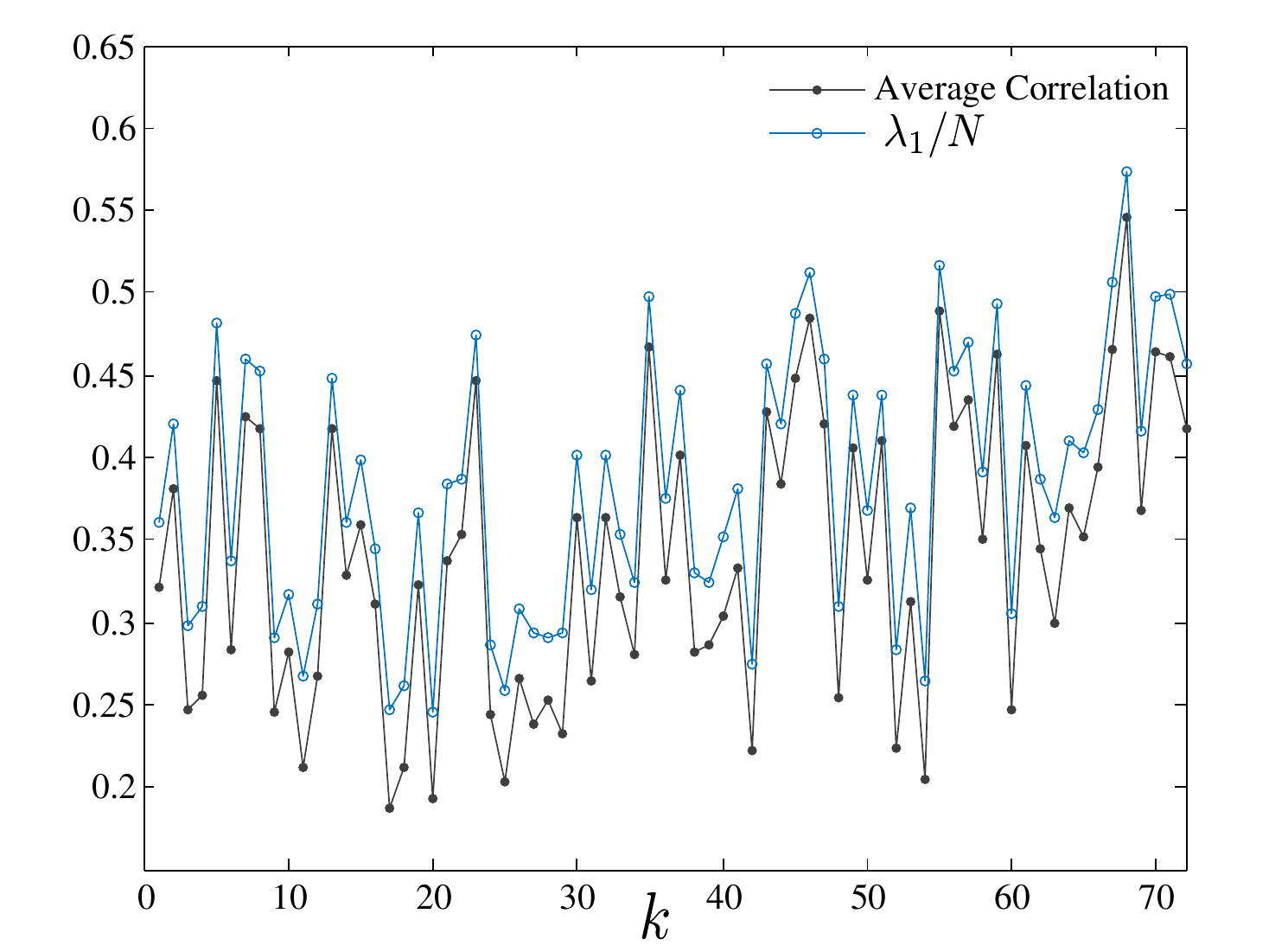}}
		\subfigure[]{\includegraphics [scale=0.4] {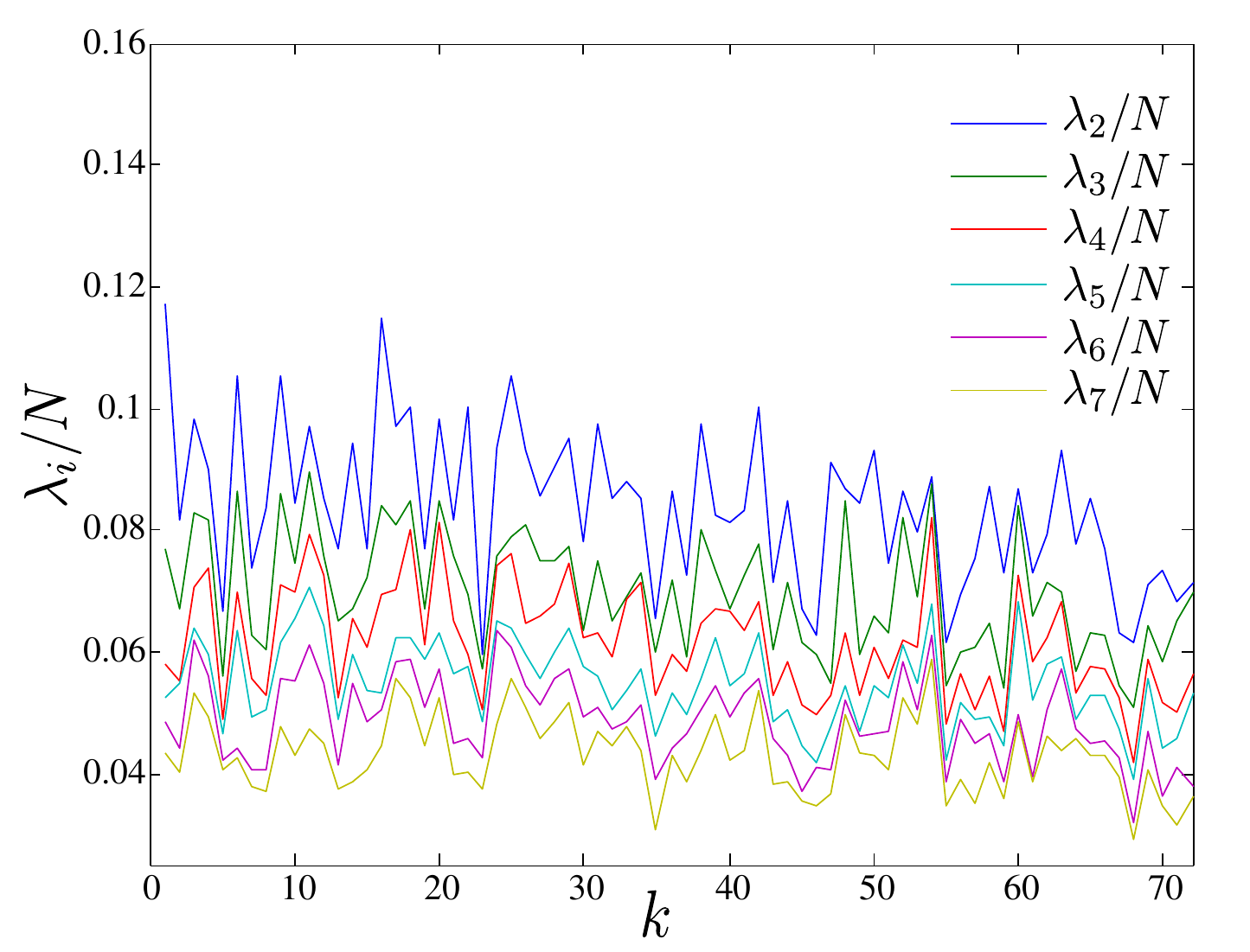}}           
    	\end{center}
    \caption{Largest eigenvalues structure for the CAC $40$, $T=5$ minute bin. (a) Average correlation between stocks (blue) and top eigenvalue $\lambda _{1}/N$ (green) of the correlation matrix $C_{\alpha \beta }(k)$. (b) Smaller eigenvalues.}
\label{fig:Fig5}
\end{figure}

In figure~\ref{fig:Fig5}(a) we show the average correlation between stocks (blue) and top eigenvalue $\lambda_{1}/N$ (green) for the CAC $40$. As can be seen the largest eigenvalue is a measure of the average correlation between stocks \cite{7, 13, 14, 15, 16}. This average correlation increases during the day from a value around $0.35$ to a value around $0.45$ when the market closes. For the case of smaller eigenvalues, what we can see is that the amplitude of risk factors decreases during the day (figure~\ref{fig:Fig5}(b)), as more and more risk is carried by the market factor (figure~\ref{fig:Fig5}(a)) \cite{7}.%

\begin{figure}[h!]
        \begin{center}
        \subfigure[$\lambda_{1}/N$]
        {\includegraphics [scale=0.4] {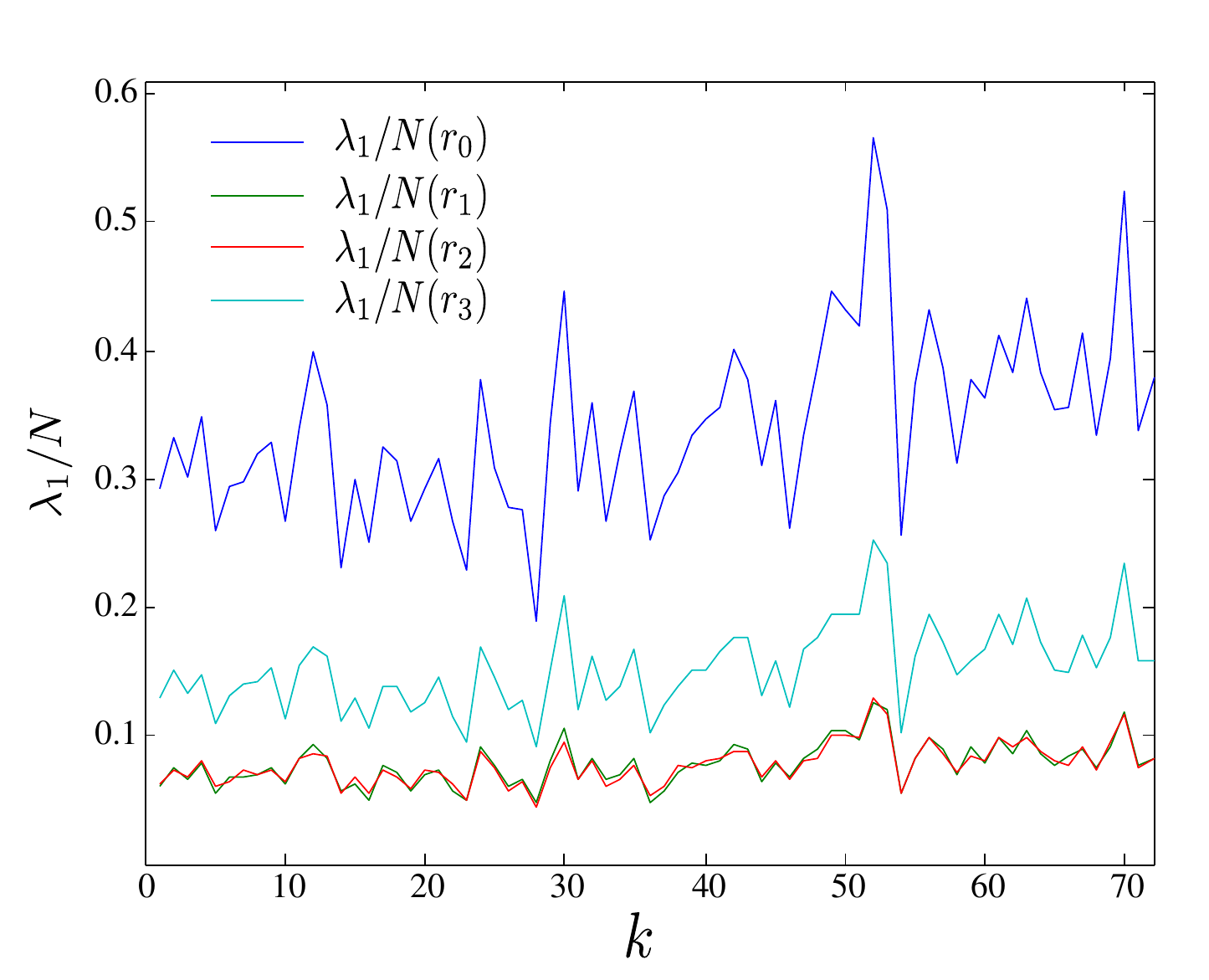}}
		\subfigure[$\lambda_{1}/N_{0}$]{\includegraphics [scale=0.4] {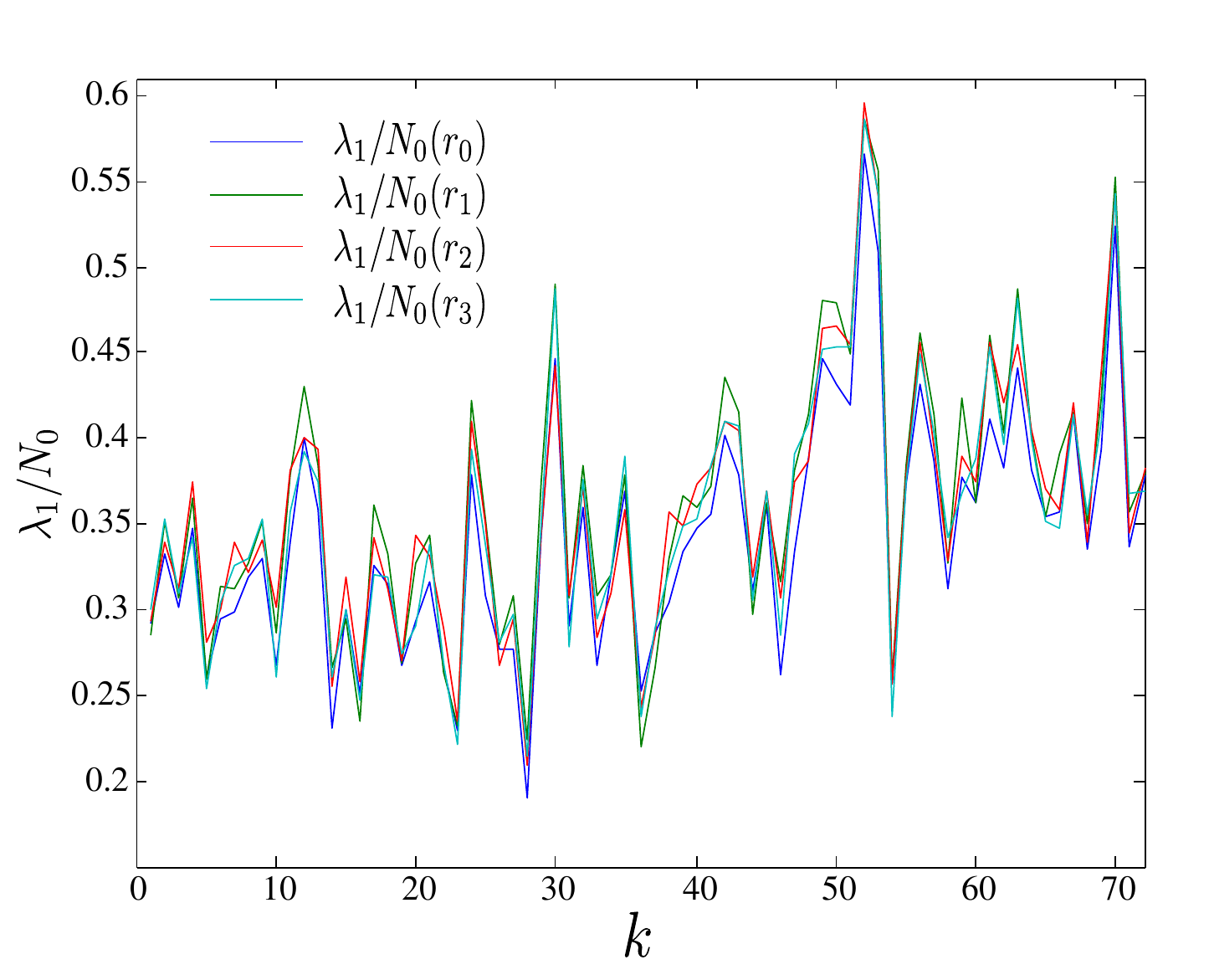}}           
    	\end{center}
    \caption{Top eigenvalue $\lambda_{1}/N$ (a) and $\lambda_{1}/N_{0}$ (b) for the S\&P $500$ for $4$ different sets of stocks: $r_{0}$ (blue), $r_{1}$ (green), $r_{2}$ (red) and $r_{3}$(clear blue). $T=5$ minute bin.}
\label{fig:Fig6}
\end{figure}

In order to simplify the computation of the $N^{2}$ correlation matrices for each bin $k$ in the case of the S\&P $500$, we computed the correlation matrix $C_{\alpha \beta }$ for $4$ different sets of stocks: $r_{0}$: composed by the $100$ first stocks of the S\&P $500$; $r_{1,2}$: composed by  $100$ stocks randomly picked; and $r_{3}$: composed by $200$ stocks randomly picked. Figure~\ref{fig:Fig6}(a) shows $\frac{\lambda _{1}}{N}$ as function of the bins. Although the values of the eigenvalues seem to be out of scale, it can be seen clearly that the average correlation increases during the day. This scale conflict is solved by normalizing the value of the top eigenvalue not by $N$ but by the sample size $N_{0}$ (i.e. $100$ or $200$) (figure~\ref{fig:Fig6}(b)). As can be seen the average correlation of the index can be computed by taking a subset of it which means that actually just the more capitalized stocks in the index drive the rest of stocks.%

\section{Intra-day Seasonalities for Relative Prices}
In this section we will report the results we found for the S\&P $500$. Similar results were found also for the CAC $40$. We will see how in the case of the relative prices these intra-day seasonalities are independent of the size of the bin, also independent of the index we consider (but characteristic for each index) however this is not the case for the returns.

\subsection{Single Stock Intra-day Seasonalities}
\begin{figure}[h!]
        \begin{center}
        \subfigure[MEAN]{\includegraphics [scale=0.4] {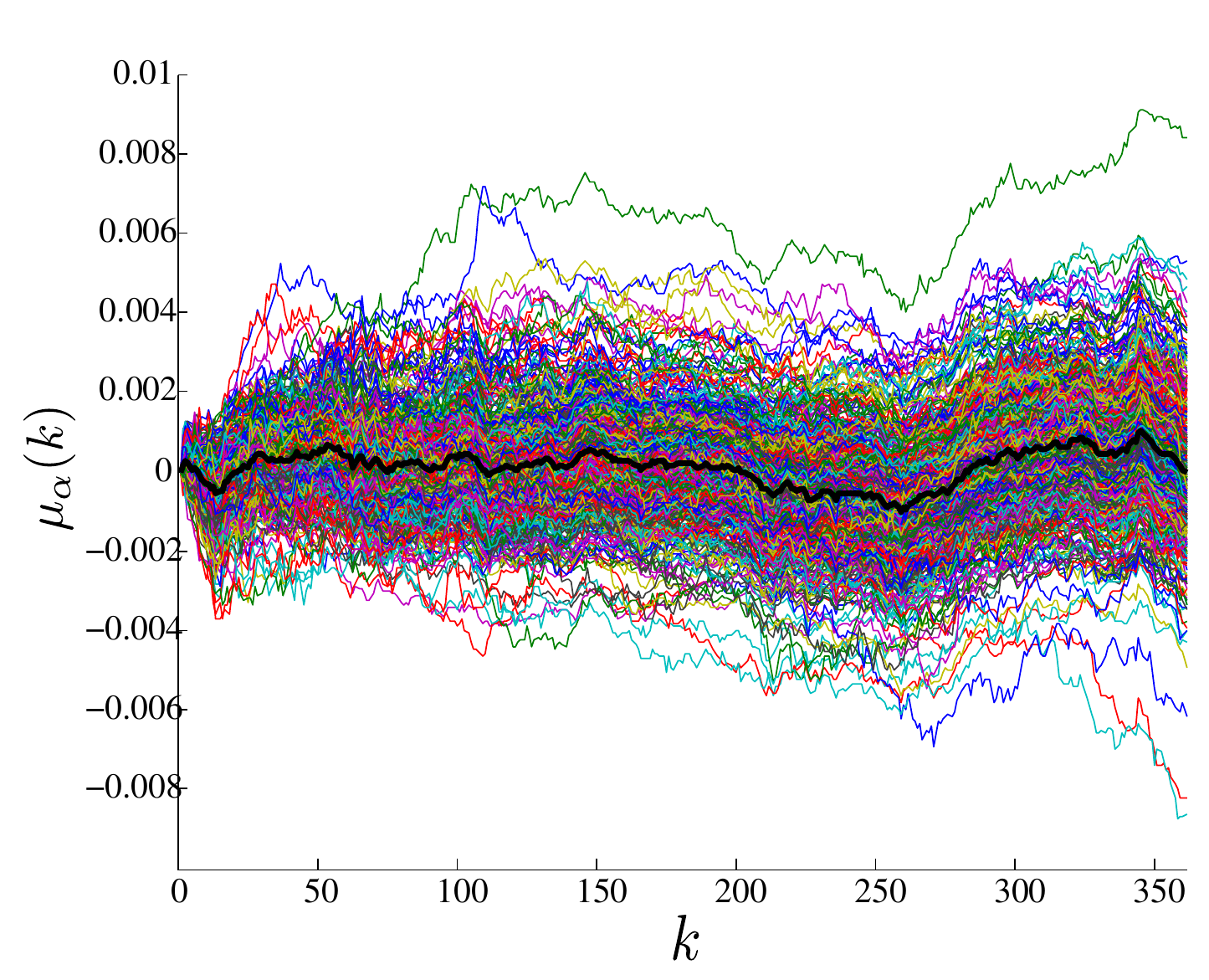}}
		\subfigure[VOLATILITY]{\includegraphics [scale=0.4] {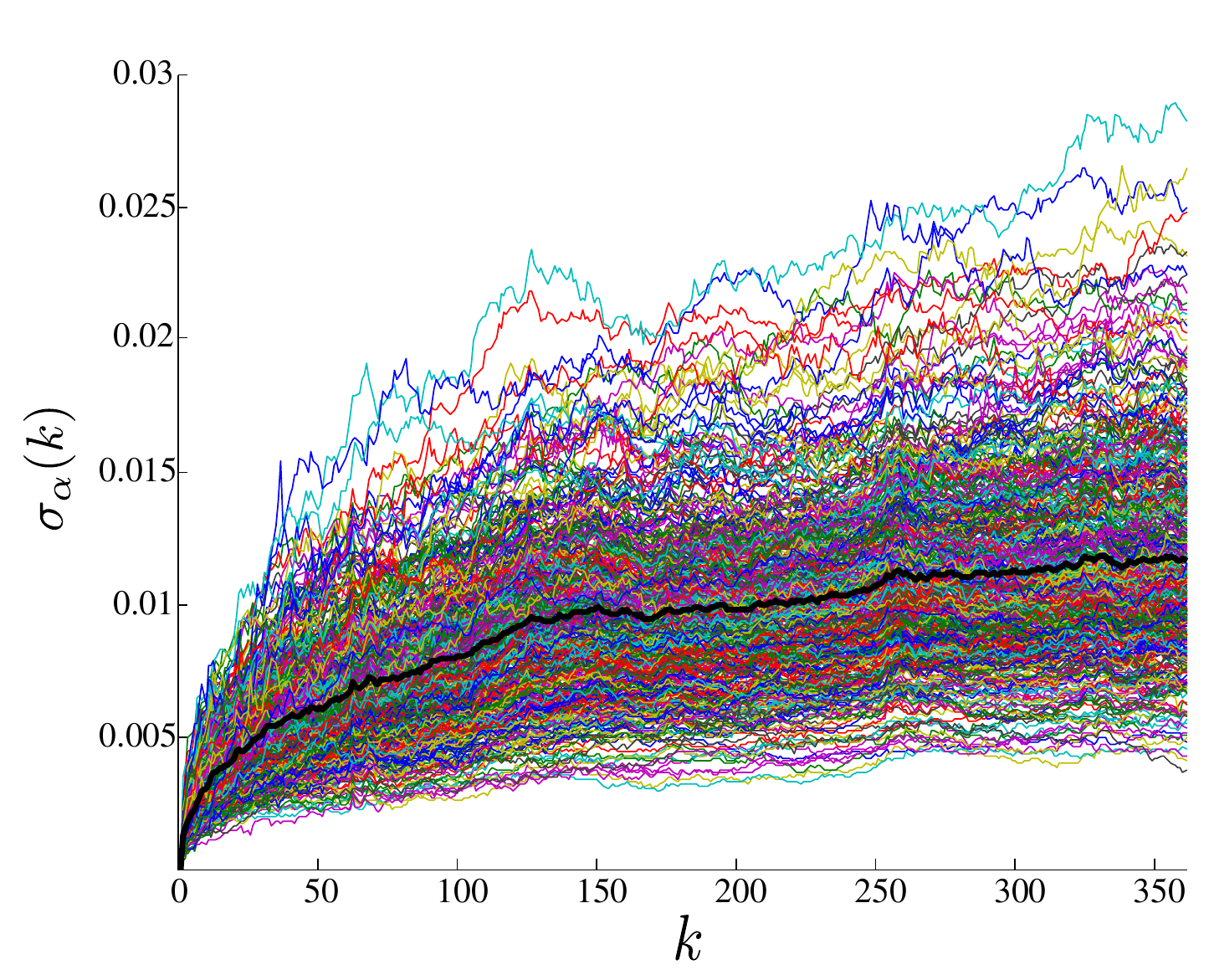}}\\ 
        \subfigure[SKEWNESS]{\includegraphics [scale=0.4] {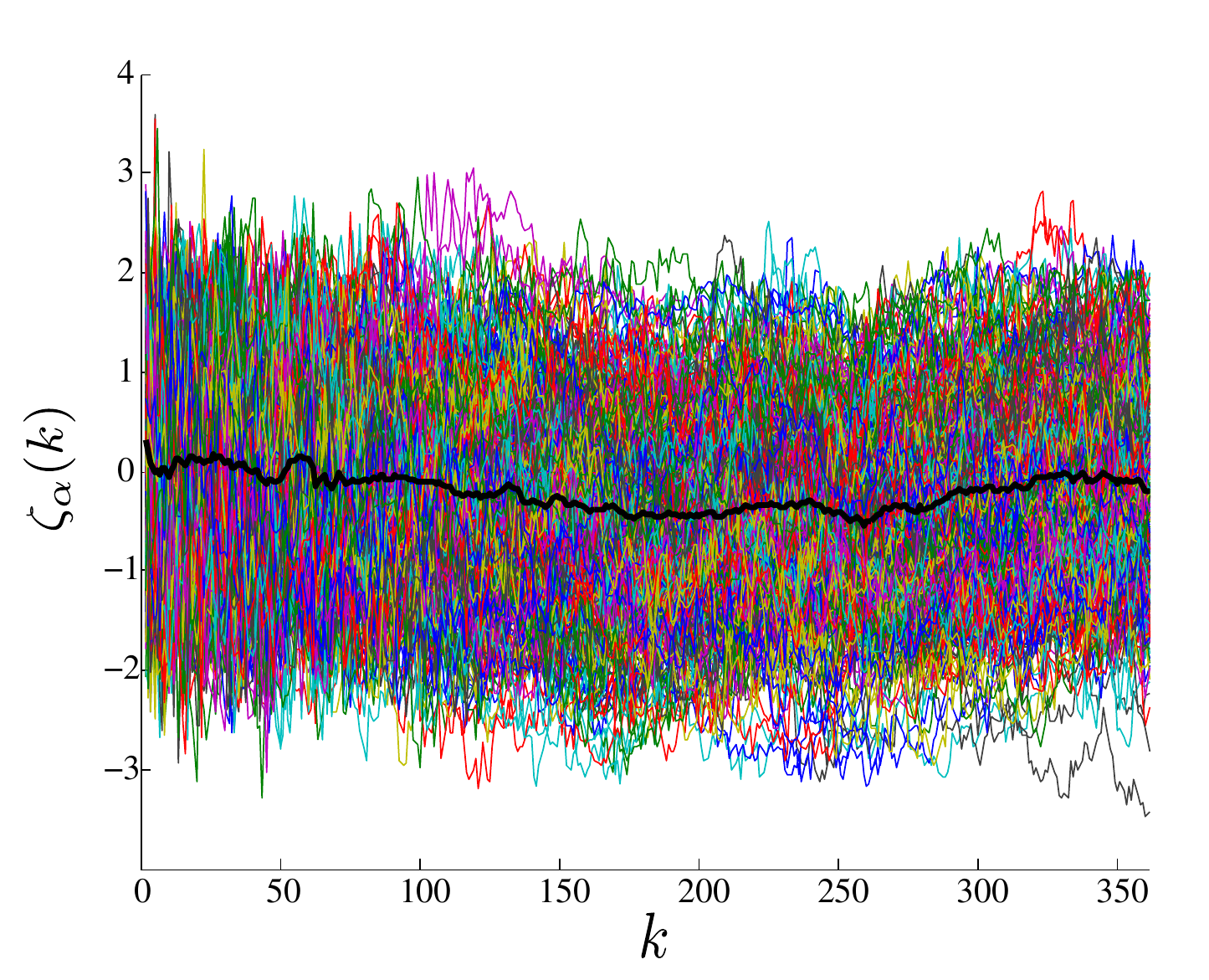}}
        \subfigure[KURTOSIS]{\includegraphics [scale=0.4] {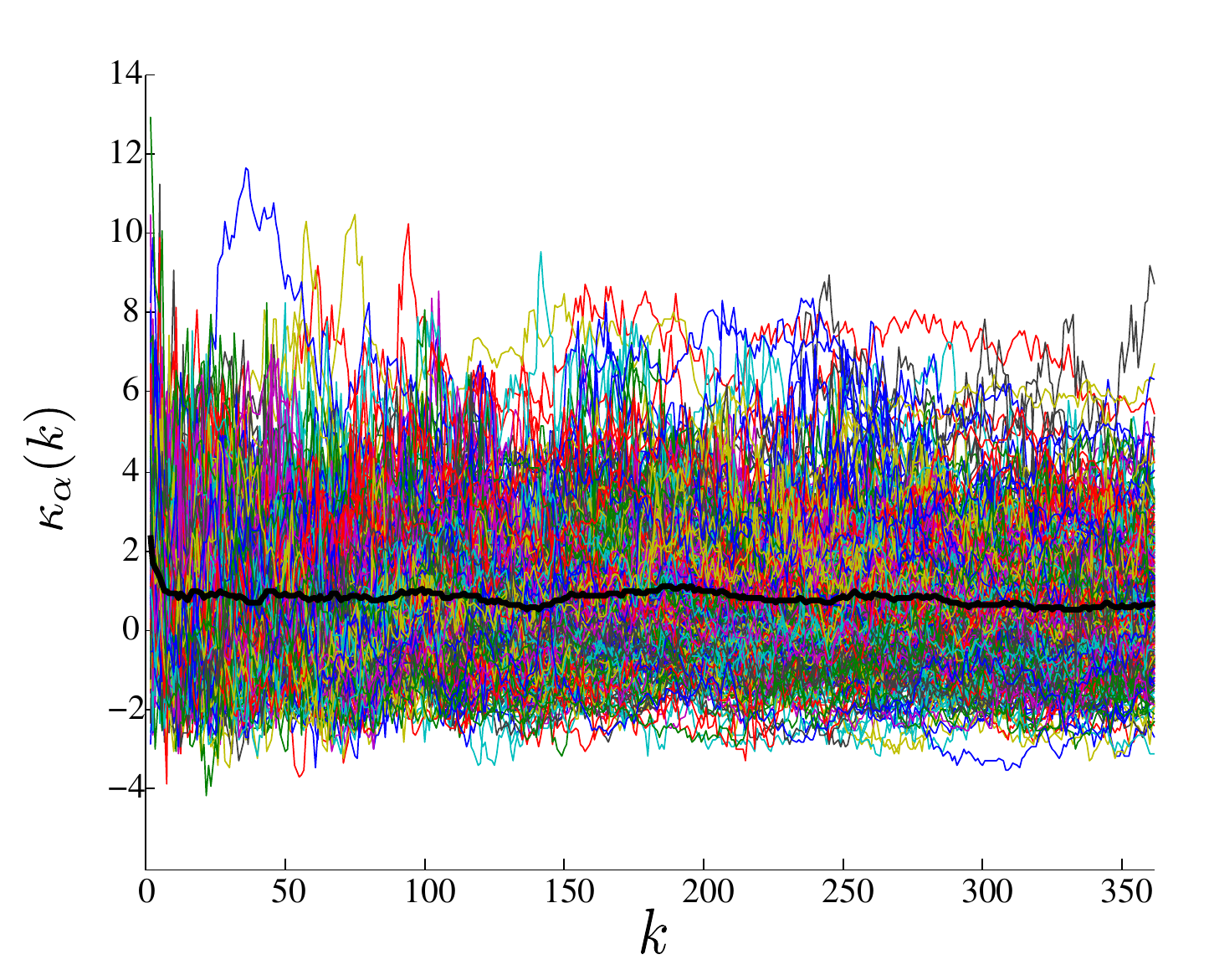}}                  
    	\end{center}
    \caption{Single Stock Intra-day Seasonalities: Stock average of the single stock mean, volatility, skewness and kurtosis for the S\&P $500$ (black). $T=1$ minute bin.}
\label{fig:Fig7}
\end{figure}

Each path in figure~\ref{fig:Fig7} represents the evolution of a particular moment of one of the stocks that compose the S\&P $500$ (i.e. one path, one stock moment). The stock average of the single stock mean $\left[\mu _{\alpha }(k)\right]$, volatility $\left[\sigma_{\alpha }(k)\right]$, skewness $\left[\zeta_{\alpha }(k)\right]$ and kurtosis $\left[\kappa_{\alpha }(k)\right]$ of the S\&P $500$ are shown in black. The stock average of the single stock mean varies around zero. The average volatility increases logarithmically with time. The skewness varies between $[-3,3]$ with an average value of zero. The single stock kurtosis takes values between $[-2,6]$ with an average value of one and its stock average starts from a value around $2$ in the very beginning of the day and decreases quickly to the mean value $1$ in the first minutes of the day.%

\subsection{Cross-Sectional Intra-day Seasonalities}
\begin{figure}[h!]
        \begin{center}
        \subfigure[MEAN]{\includegraphics [scale=0.4] {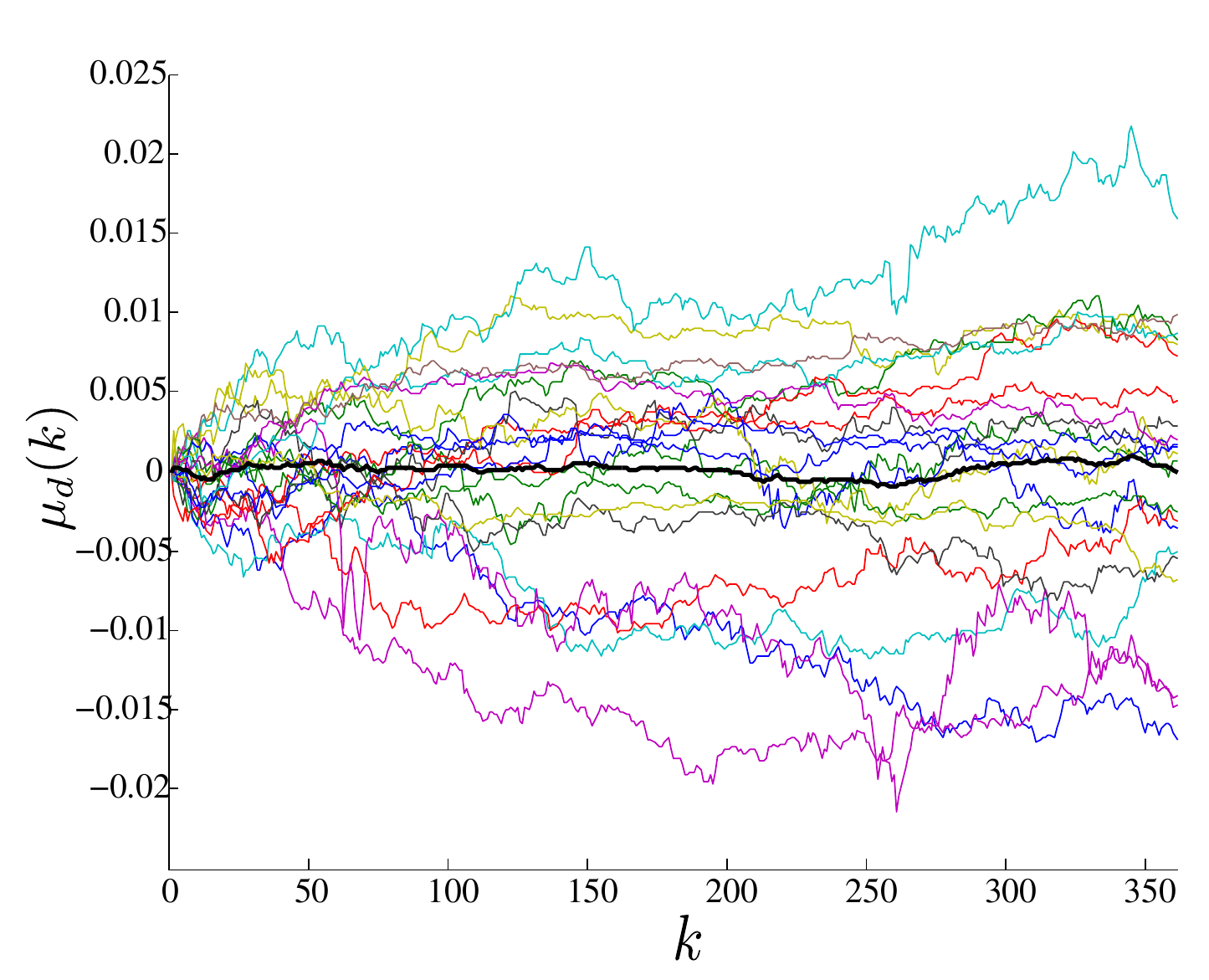}}
		\subfigure[VOLATILITY]{\includegraphics [scale=0.4] {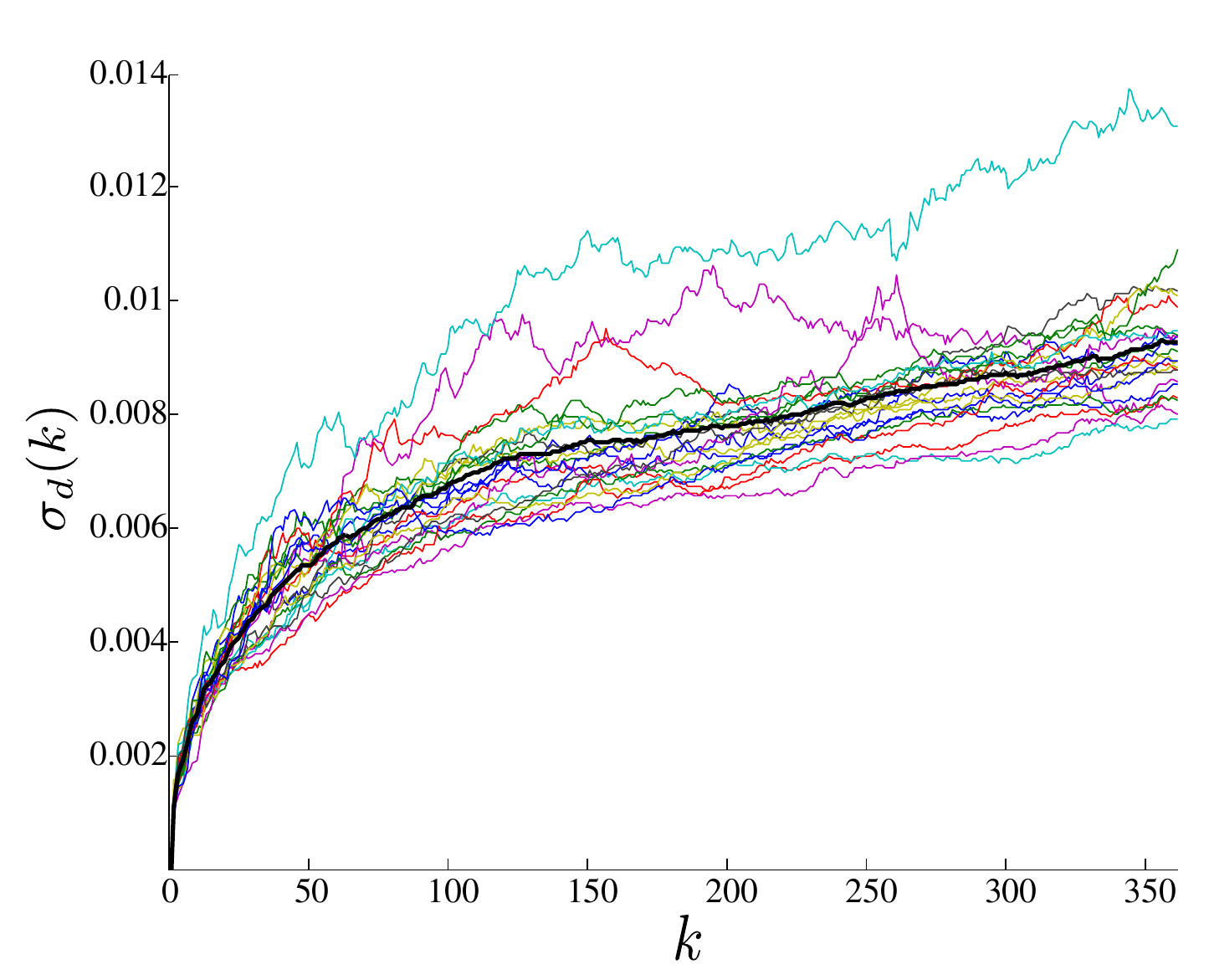}}\\ 
        \subfigure[SKEWNESS]{\includegraphics [scale=0.4] {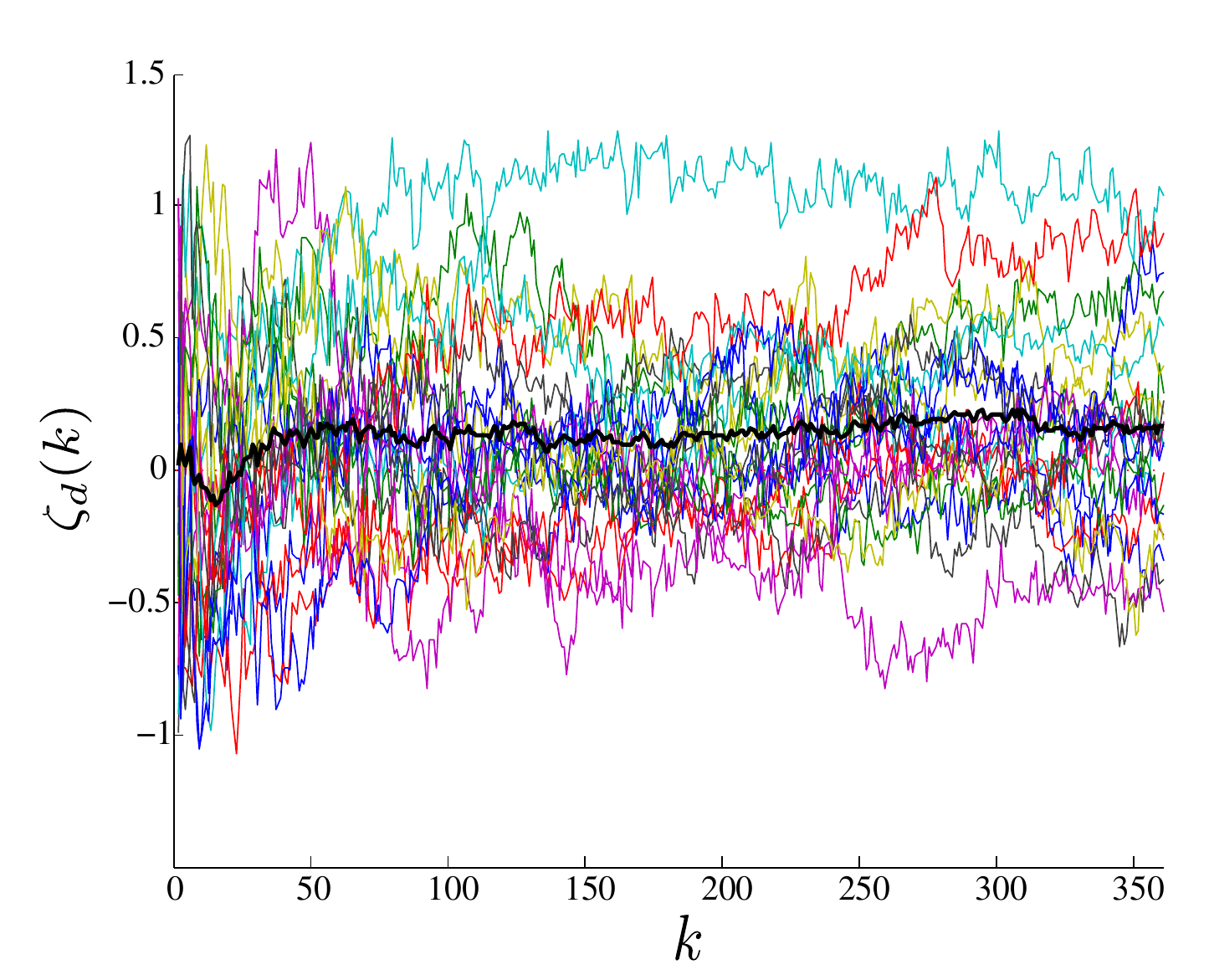}}
        \subfigure[KURTOSIS]{\includegraphics [scale=0.4] {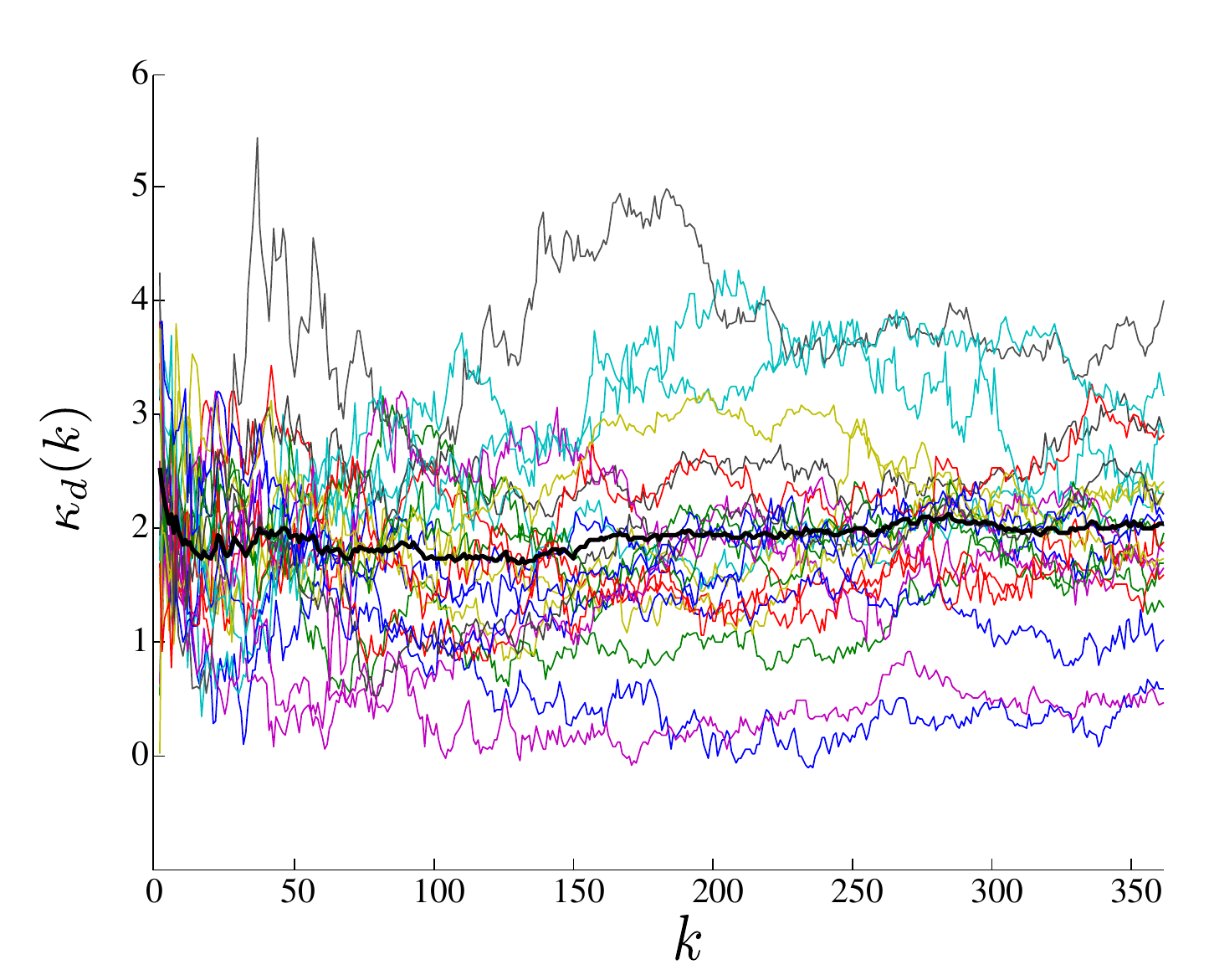}}                  
    	\end{center}
    \caption{Cross-Sectional Intra-day Seasonalities: Time average of the cross-sectional mean, volatility, skewness and kurtosis for the S\&P $500$ (black). $T=1$ minute bin.}
\label{fig:Fig8}
\end{figure}

Each path in figure~\ref{fig:Fig8} represents the evolution of a particular index moment during a particular day (i.e. one path, one day moment). As in the case of the single stock volatility, the cross-sectional dispersion $\langle\sigma_{d}(k)\rangle$ increases logarithmically with respect to the time (figure~\ref{fig:Fig8}(b)). The cross-sectional skewness $\langle\zeta_{d}(k)\rangle$ takes values in the interval $[-1,1]$ with an average value of zero (figure~\ref{fig:Fig8}(c)). The average kurtosis $\langle\kappa _{d}(k)\rangle$ starts from a value around $2.5$ in the very beginning of the day and decreases quickly to the mean value $2$ in the first minutes of the day (figure~\ref{fig:Fig8}(d)).

\subsection{C-Pattern Volatilities}
\begin{figure}[h!]
\centering
\includegraphics [scale = 0.65] {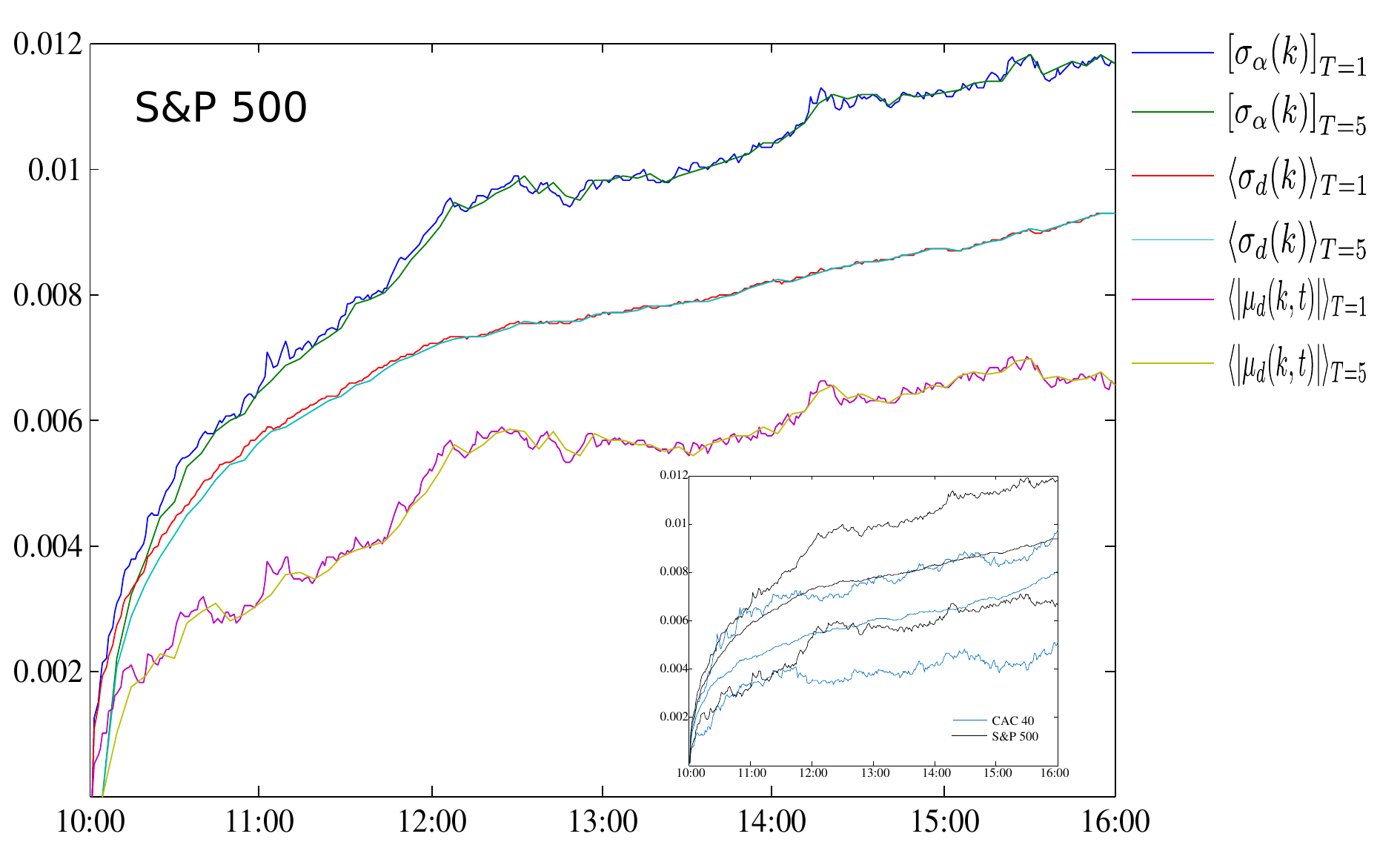}
\caption{C-Pattern Volatilities: Stock average of the single stock volatility $\left[\sigma _{\alpha }(k)\right]$, time average of the cross-sectional volatility $\langle\sigma_{d}(k,t)\rangle$ and the average absolute value of the cross-sectional mean $\langle|\mu_{d}|\rangle$ for the relative prices of the S\&P $500$. $T=1$ and $T=5$ minute bin. Inset: CAC $40$ (blue) and S\&P $500$ (black). }
\label{fig:Fig9}
\end{figure}

Similarly as we did in section 3.3 for returns, in figure~\ref{fig:Fig9} we show a comparative plot between the stock average of the single stock volatility $\left[\sigma _{\alpha }(k)\right]$, the time average of the cross-sectional volatility $\langle\sigma _{d}(k,t)\rangle$ and the average absolute value of the cross-sectional mean $\langle|\mu _{d}|\rangle$ for the relative prices
of the S\&P $500$, and for $T=1$ and $T=5$ minute bin. As can be seen, these three measures exhibit the same kind of intra-day pattern (as it did in the case of the returns). But the most important fact is to notice that this intra-day seasonality is independent of the size of the bin, also independent of the index we consider, but characteristic for each index (see inset figure~\ref{fig:Fig9}). 

\section{Intra-day Patterns and Bin Size}
\begin{figure}[h!]
        \begin{center}
        		\subfigure[$T = 0.5$ minute bin]{\includegraphics [scale=0.35] {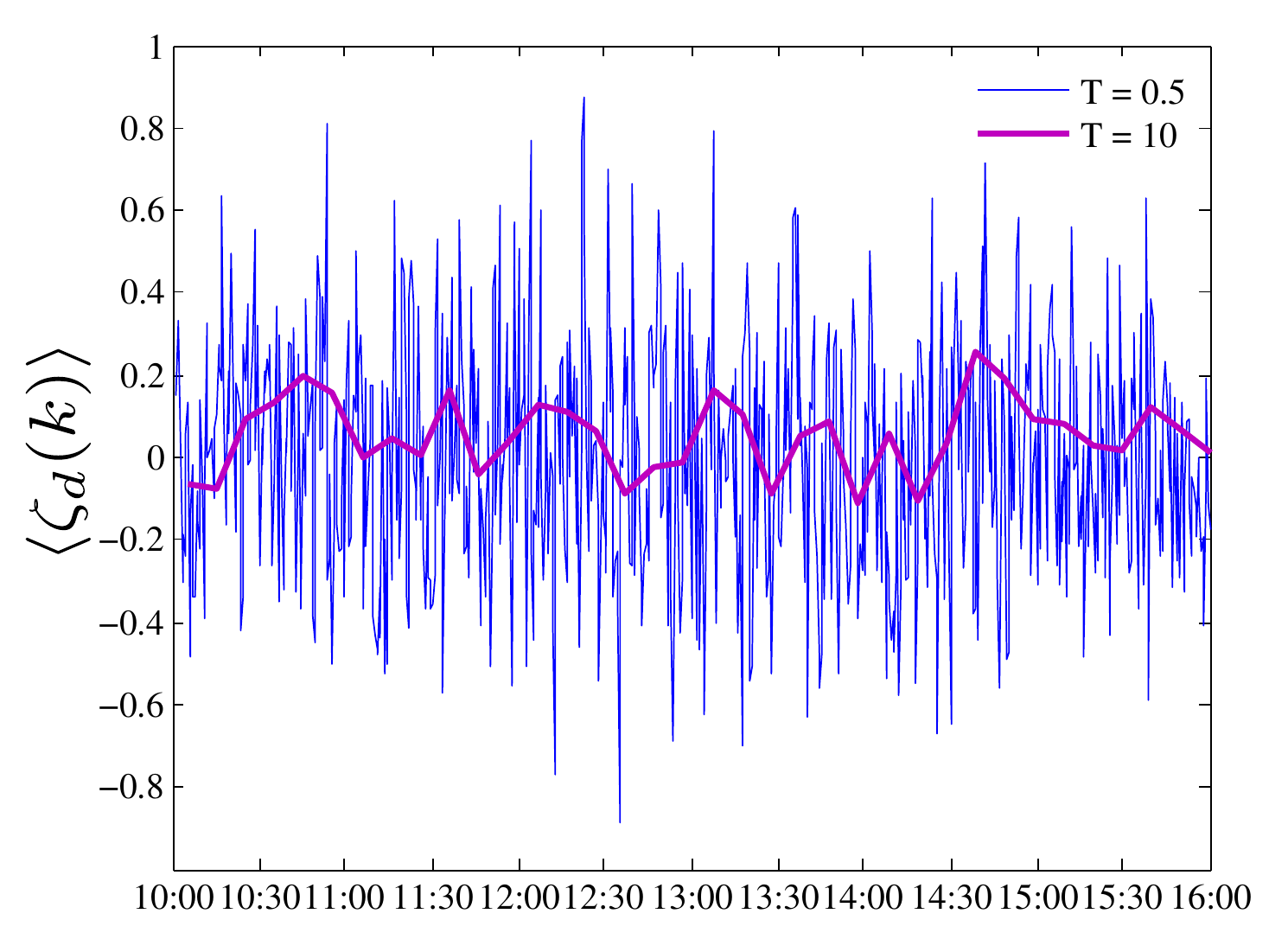}}
			\subfigure[$T = 1$ minute bin]{\includegraphics [scale=0.35] {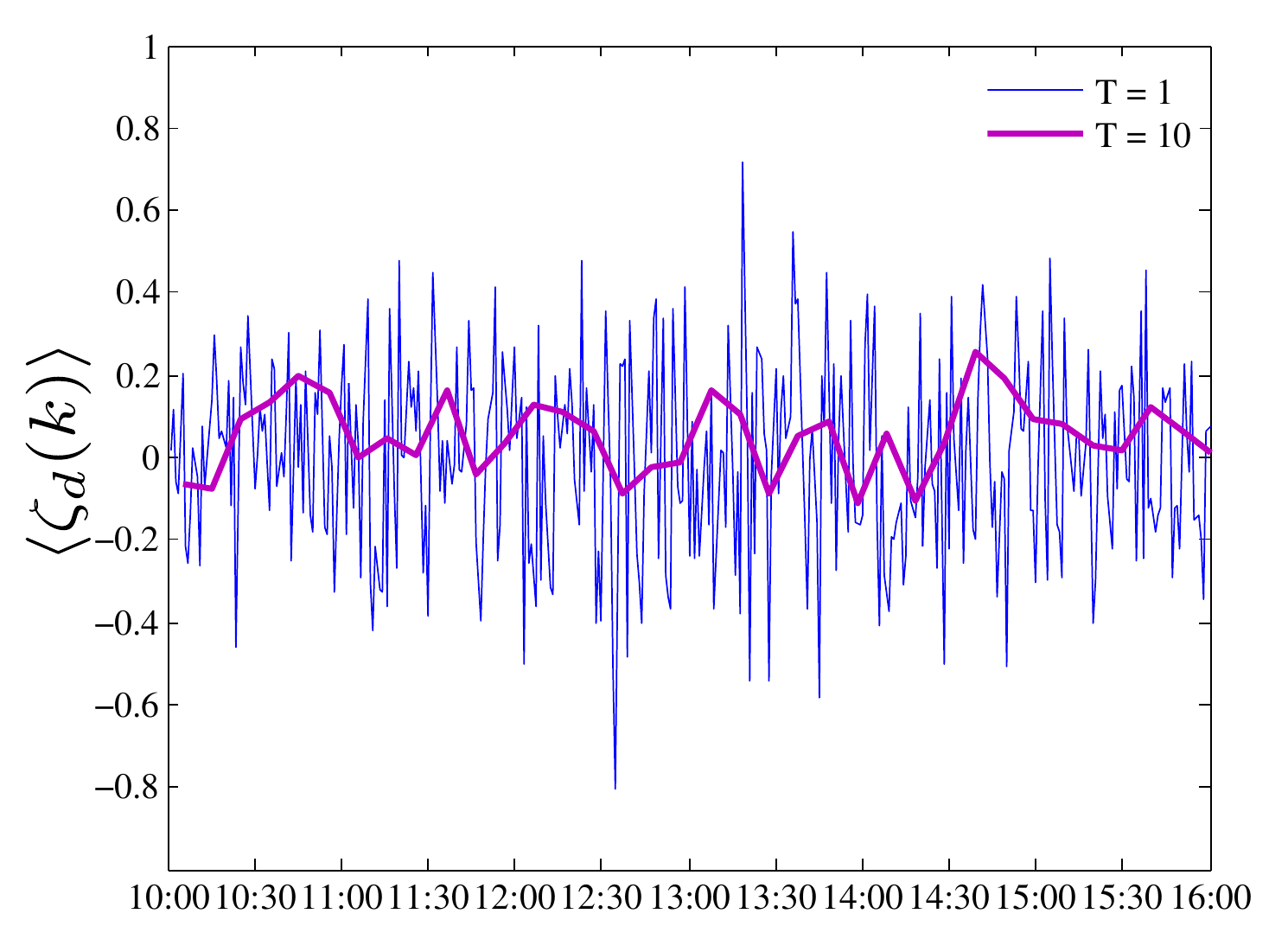}}                 
    		\end{center}
    		\caption{Time average of the cross-sectional skewness: Comparison of the intra-day patterns for $T=0.5$ and $T=1$ against $T = 10$ minute bin for the S\&P $500$.}
\label{fig:Fig10}
\end{figure}

As we saw in the last section, the volatilities for the relative prices
exhibit the same kind of intra-day pattern (figure~\ref{fig:Fig9}). This intra-day seasonality is independent of the size of the bin, and the index we consider, but characteristic for each index. Actually, this is not true in the case of the returns as we already suggested in section 3.3 from figure~\ref{fig:Fig4} . If we consider the odd moments (mean and skewness) of the returns, the behaviour is basically the same (noisy around zero) and without any particular pattern, independently of the bin size (as can be seen in figures~\ref{fig:Fig2},~\ref{fig:Fig3} and~\ref{fig:Fig10}). 

\begin{figure}[h!]
\centering
\includegraphics [scale=0.7] {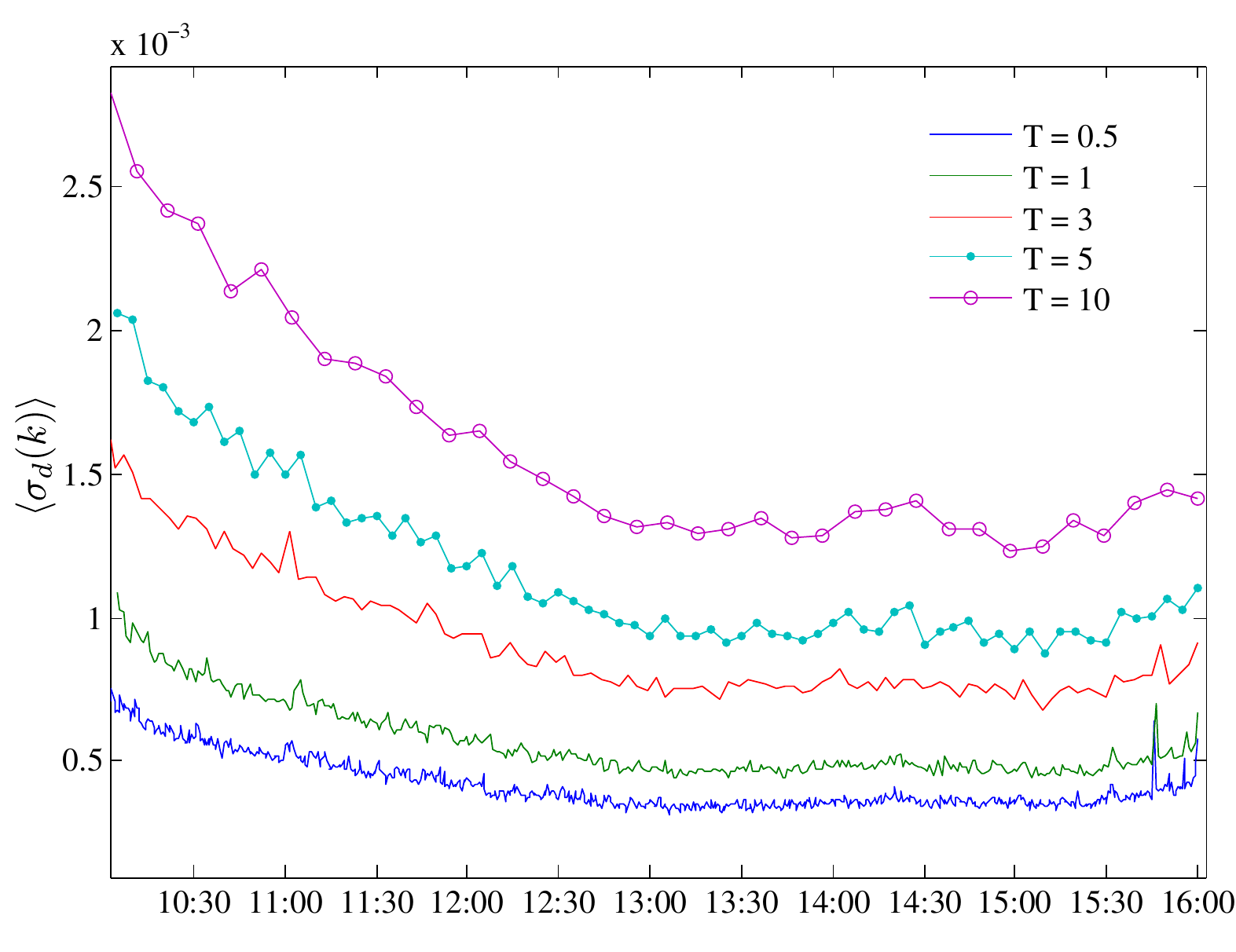}
\caption{Bin size dependence in the U-pattern volatilities: Time average of the cross-sectional volatility for the S\&P $500$ for $5$ different values of bin size $T$.}
\label{fig:Fig11}
\end{figure}

\begin{figure}[h!]
\centering
\includegraphics [scale=0.7] {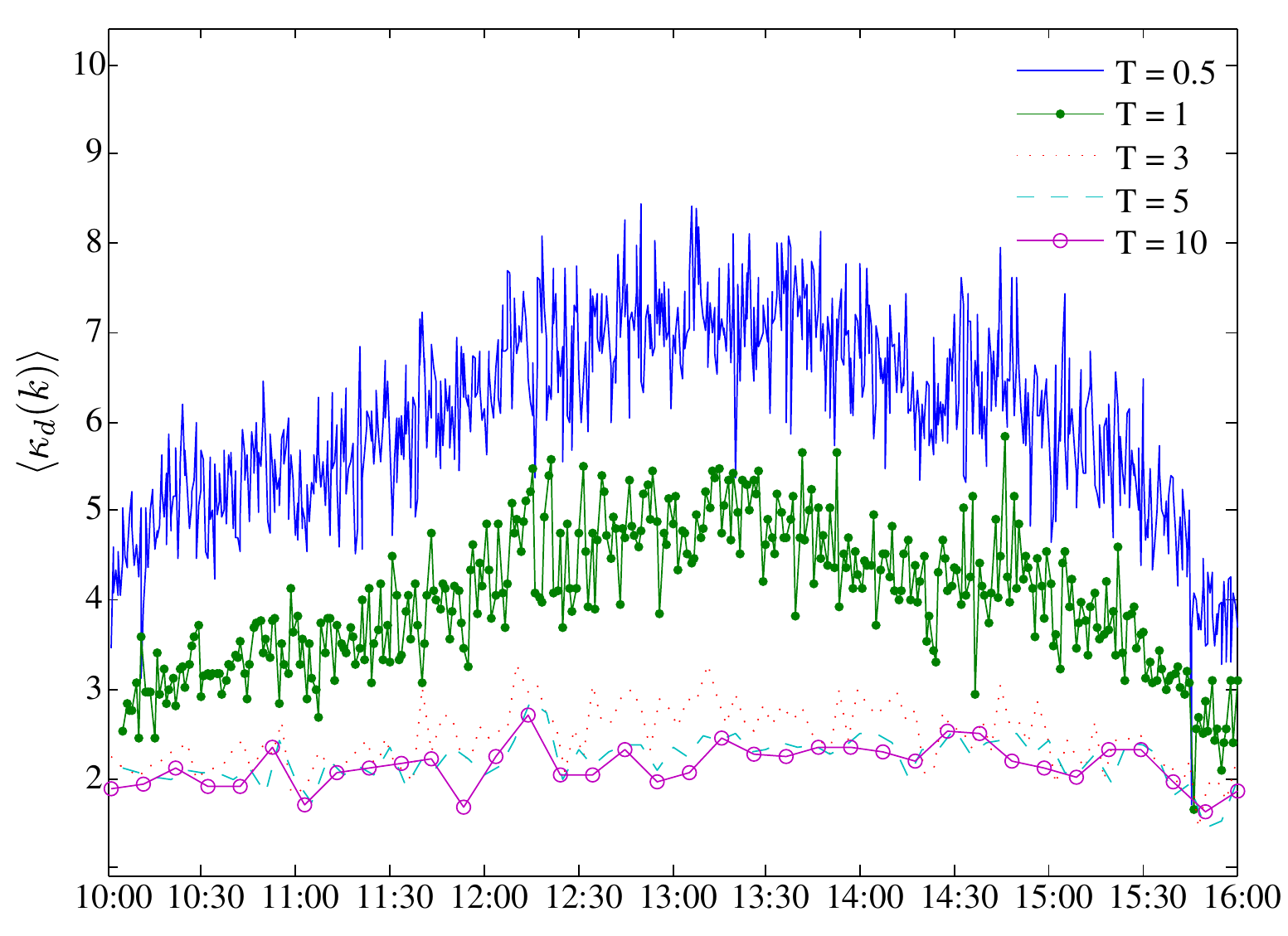}
\caption{Bin size dependence in the inverted U-pattern kurtosis: Time average of the cross-sectional kurtosis for the S\&P $500$ for $5$ different values of bin size $T$.}
\label{fig:Fig12}
\end{figure}

But for the case of the even moments of returns,
although they exhibit the well known U and inverted U-patterns, these patterns depend on the bin size. This fact is well illustrated through figures~\ref{fig:Fig11} and~\ref{fig:Fig12} in where we have chosen $5$ different values of bin size from $T=0.5$ to $T=10$ minutes. In these figures we show the time average of the cross-sectional volatility and kurtosis for the S\&P $500$  but a similar bin size dependence can be shown for the CAC $40$ or any other index and also for the time average of the single stock volatility and kurtosis.

By other hand the kurtosis is a  decreasing function of the size of the bin and the inverted U-pattern is evident just when we consider ``small'' bin sizes, in our case this occurs for $T=1$ and $T=0.5$ minute bin (figure~\ref{fig:Fig12}). This represents a confirmation that on small scales the returns have heavier tails, and on long time scales they are more Gaussian \cite{4, 8, 9, 10}.

\section{Intra-day Abnormal Patterns}
One of the motivations we had in order to explore into the intra-day seasonalities for relative prices was due to Kaisoji's previous work \cite{12}. In his work he found that the upper tail of the complementary cumulative
distribution function of the ensemble of the relative prices in the high value of the price is well described by a power-law distribution which when its exponent approached two, the Japan's internet bubble burst.
Taking into consideration our recent findings we suggest the use of the bin size independence for intra-day patterns in relative prices in order to characterize ``atypical days'' for indexes and ``anomalous behaviours'' for stocks. 

\begin{figure}[h!]
        \begin{center}
        \subfigure[MEAN]{\includegraphics [scale=0.4] {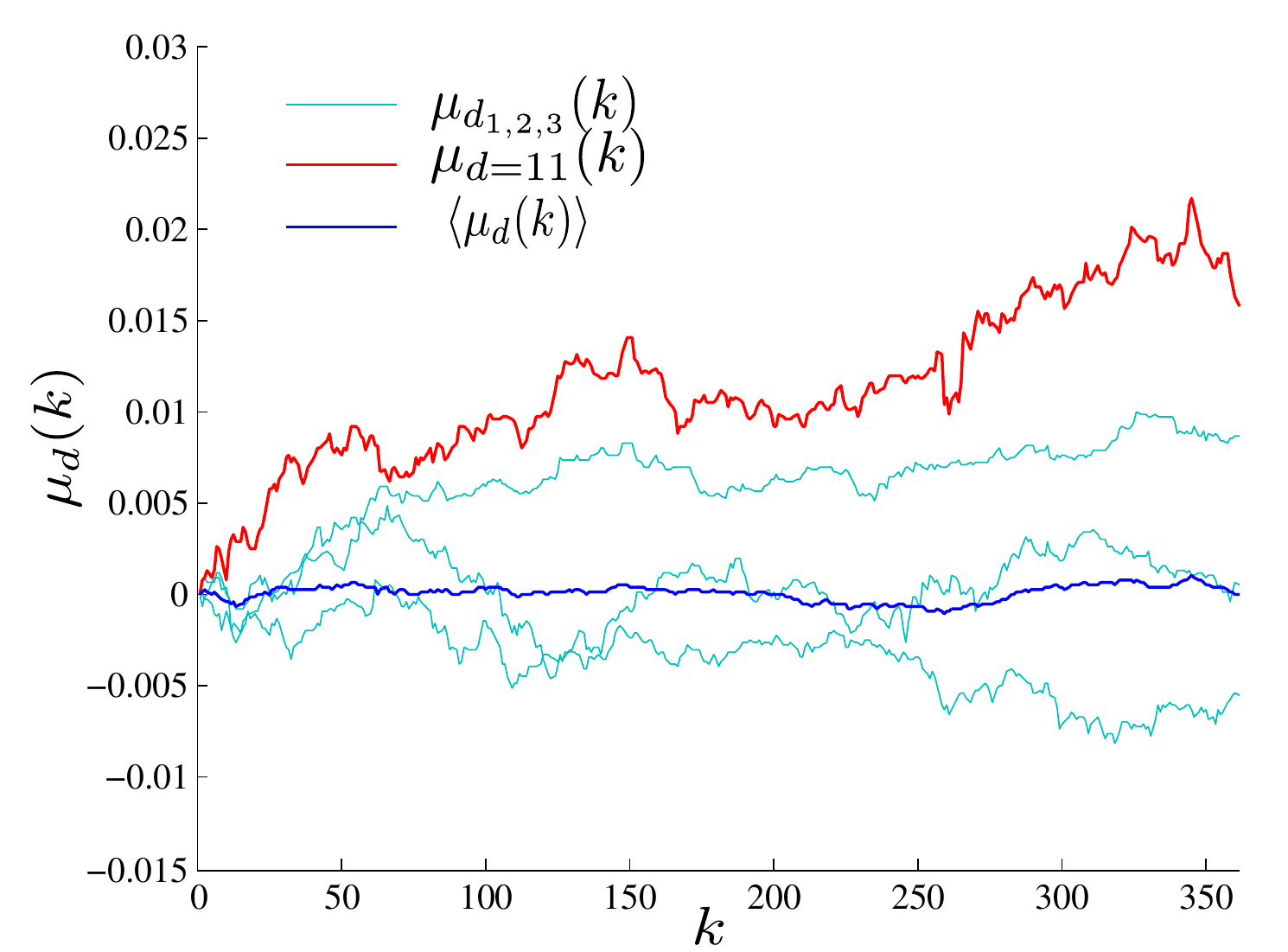}}
        \subfigure[VOLATILITY]{\includegraphics [scale=0.4] {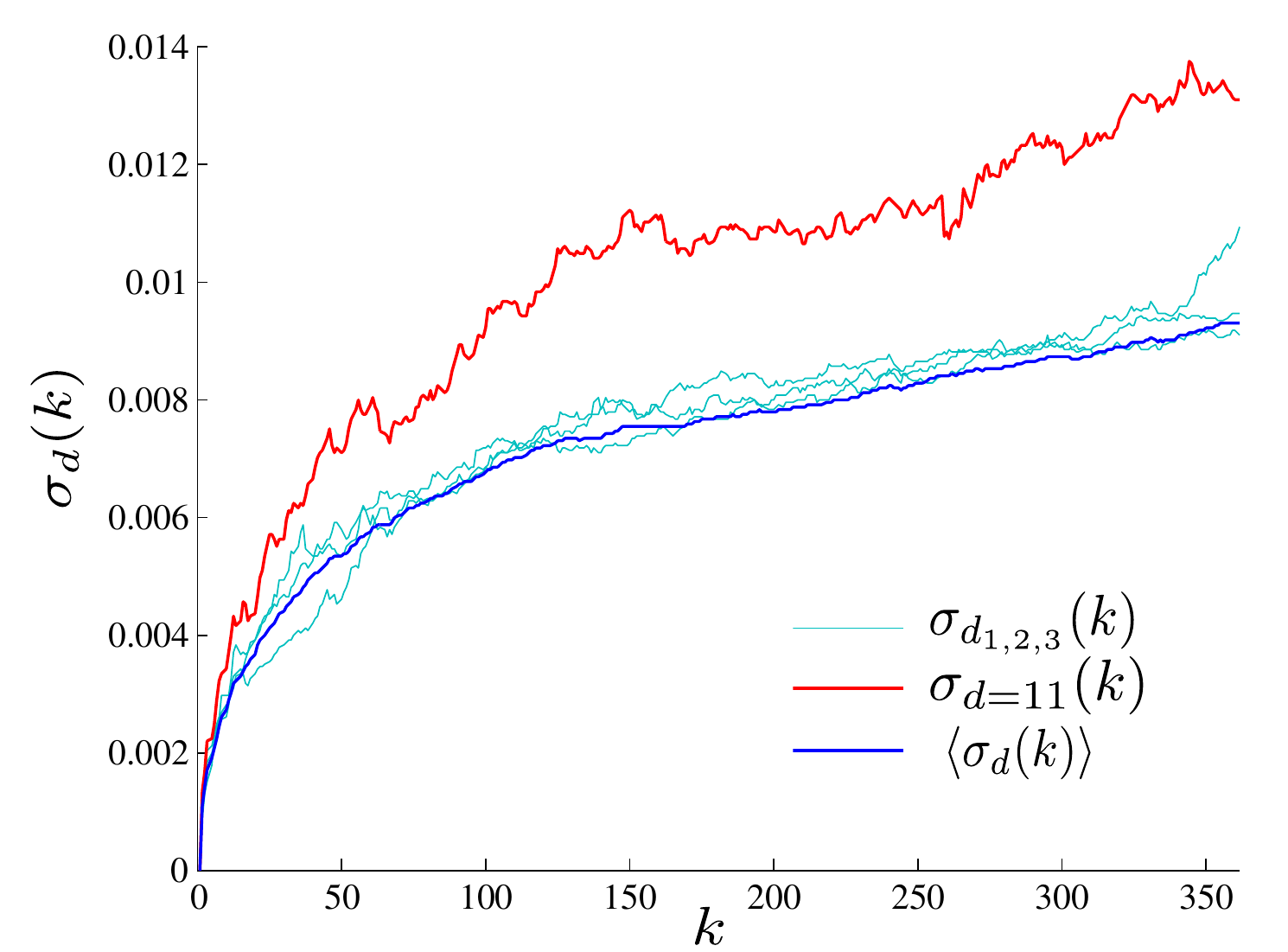}}                  
    	\end{center}
    \caption{S\&P $500$ Atypical Day: Time average of the cross-sectional mean and volatility (blue), cross-sectional mean and volatility of the S\&P $500$ during day $11$ (red) and during three days chosen at random (clear blue).}
\label{fig:Fig13}
\end{figure}

The time average of the cross-sectional moments represents the average behaviour of a particular index moment during an average day. In figure~\ref{fig:Fig8} we showed how each path represents the evolution of a particular index moment for one of the days of the period under analysis (i.e. one path, one day moment). If we look directly into the prices of the CAC $40$ and S\&P $500$, we can observe during day $11$ a fall of the prices of the stocks that compose both indexes. During the days before and following day $11$, the (index) moments move along our intra-day pattern. Moreover, if we pick randomly one day from our period of analysis, in most of the cases our index during that day will behave as our intra-day seasonality (as in figure~\ref{fig:Fig13}), but the one for day $11$ will not. In figure~\ref{fig:Fig13} we show our (cross-sectional) intra-day seasonalities  for the mean (a) and volatility (b) in blue and in clear blue the respective cross-sectional stock moments for 3 days randomly picked. The average behaviour (of the moments) of our index during these days moves along with our intra-day pattern. This is not the case of the curve corresponding to the day $11$ shown in red which clearly diverges from the expected behaviour. This is what could be called as an ``atypical day'' for the S\&P $500$.

\begin{figure}[h!]
        \begin{center}
        \subfigure[MEAN]{\includegraphics [scale=0.4] {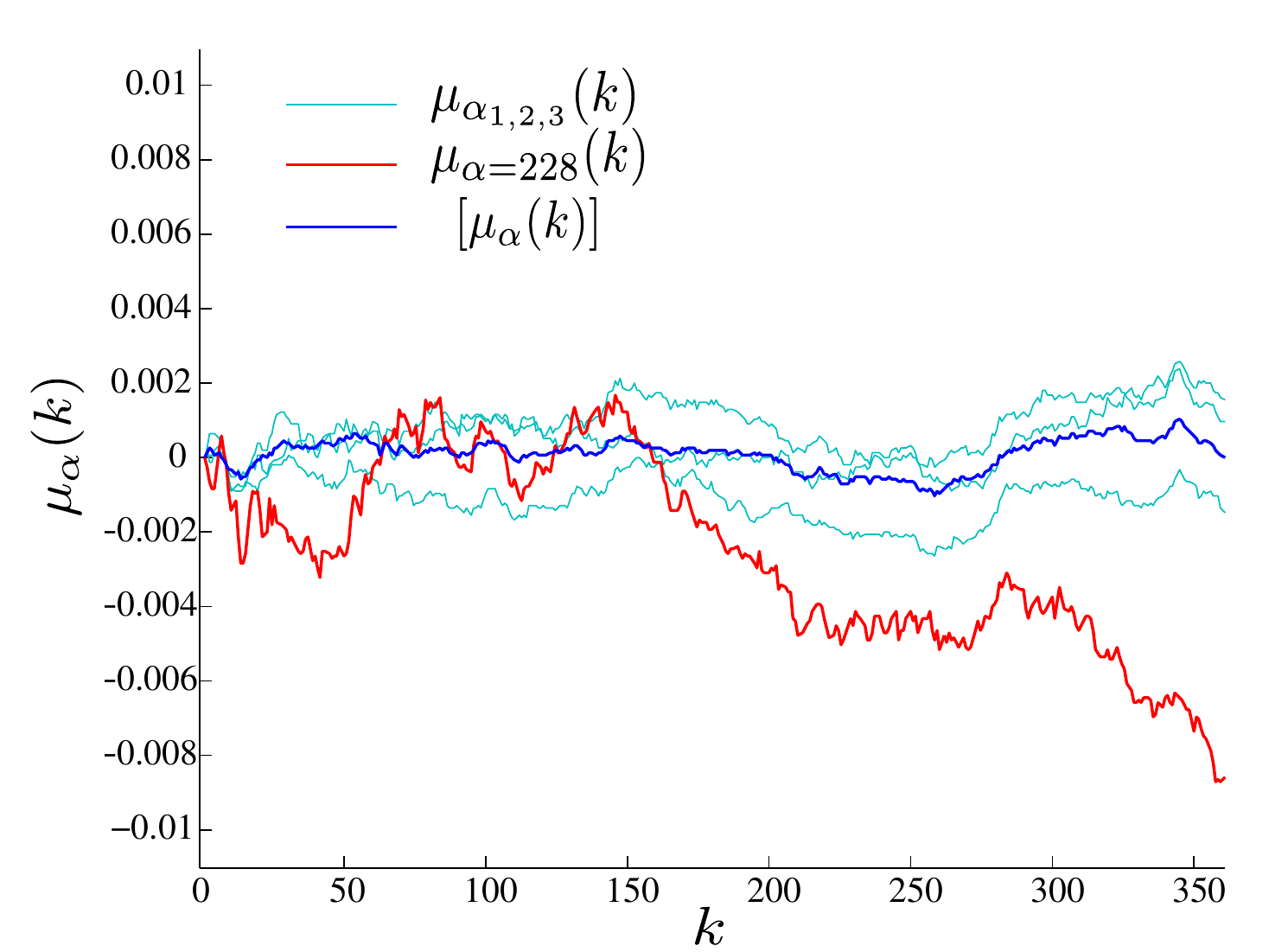}}
		\subfigure[VOLATILITY]{\includegraphics [scale=0.4] {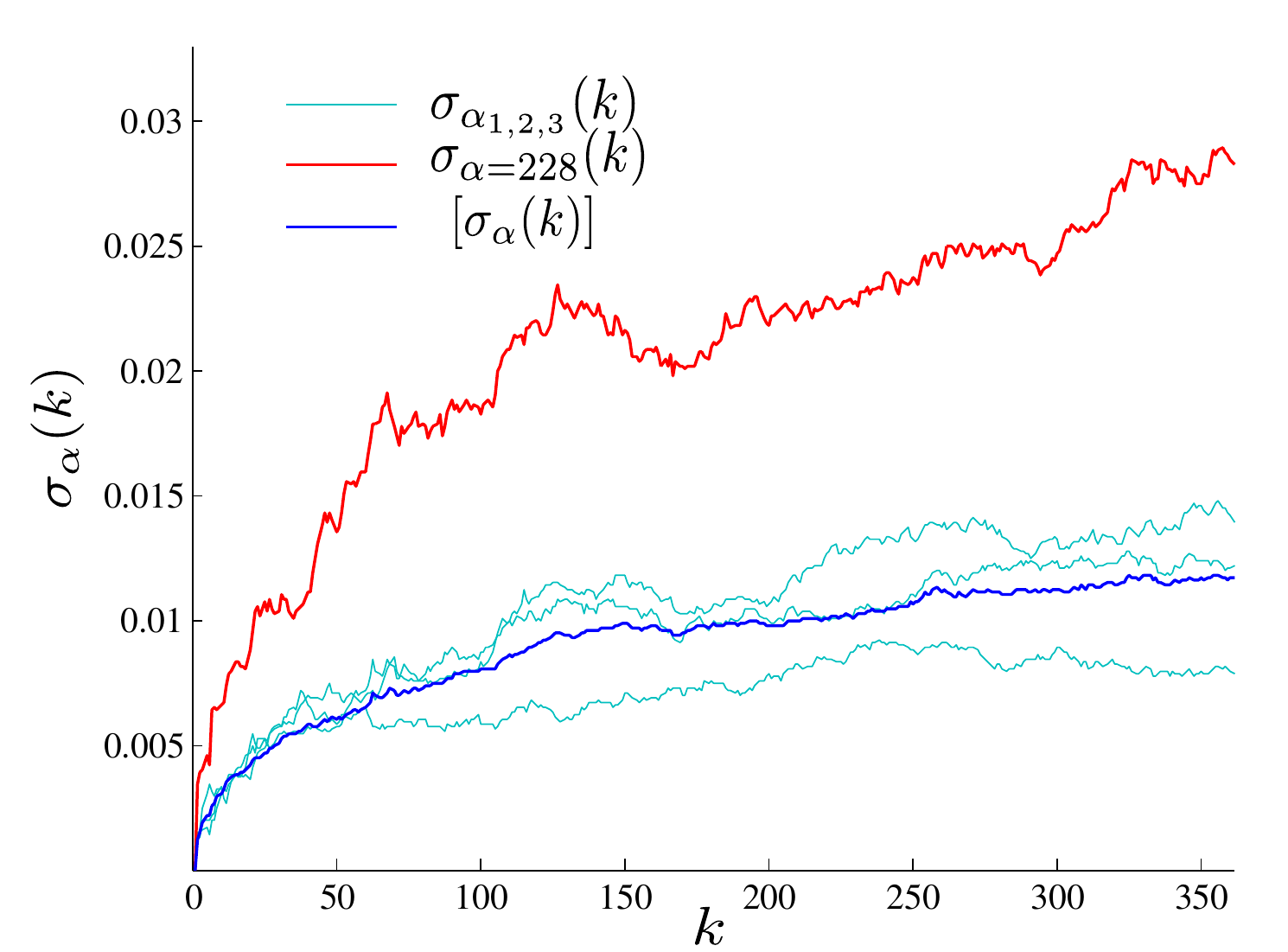}}\\ 
    	\end{center}
    \caption{Anomalous Stock Behaviour: Stock average of the single stock mean and volatility (blue), single stock mean and volatility of stock $228$ (red) and from three stocks chosen at random from the S\&P $500$ (clear blue).}
\label{fig:Fig14}
\end{figure}

We could used the same reasoning as before in order to characterize ``anomalous behaviours'' in stocks. As we said in section 4.1, each path in figure~\ref{fig:Fig7} represents the average evolution of a particular moment of one of the stocks that compose the S\&P $500$. The stock average of those single stock moments represents the average behaviour of that moment for an average stock during an average day of our period of analysis. 
That means that if from our set of stocks we pick randomly one stock, in most of the cases (its moments) will behave as our intra-day seasonality. This is clearly illustrated in figure~\ref{fig:Fig14} where we show our intra-day seasonalities  for the mean (a) and volatility (b) in blue and the respective single stock moments for 3 stocks randomly picked in clear blue. As can be seen, the average behaviour of the moments of these stocks move along  with our intra-day patterns. However this is not the case for the curves shown in red which have been chosen on  purpose to illustrate how in this case the stock $228$ behaves in an anomalous way with respect to what is expected.

\section{Discussion}
In the present report, we have analysed the intra-day seasonalities of the single and cross-sectional (or collective) stock dynamics. In order to do this, we characterized the
dynamics of a stock (or a set of stocks) by the evolution of the moments of
its returns (and relative prices) during a typical day. What we have
called ``single stock intra-day seasonalities'' is the average behaviour of the
moments of the returns (and relative prices) of an average stock in an
average day. In the same way, the cross-sectional intra-day seasonality is
not more than the average day behaviour of an index moment. We presented these intra-day seasonalities for returns (figures~\ref{fig:Fig2} and~\ref{fig:Fig3}) and relative prices (figures~\ref{fig:Fig7} and~\ref{fig:Fig8}) and compared the stock average of single stock volatility $\left[\sigma _{\alpha }(k)\right]$, the time average of the cross-sectional volatility $\langle\sigma _{d}(k,t)\rangle$ and the average absolute value of the equi-weighted index $\langle|\mu _{d}|\rangle$ (figures~\ref{fig:Fig4} and~\ref{fig:Fig9}). 

One thing that results interesting to observe, in the case of the returns, is that these ``patterns'' actually depend on the size of the bin. This fact was well illustrated with $5$ different values of bin size through figure~\ref{fig:Fig11} for volatilities and figure~\ref{fig:Fig12} for kurtosis in which its inverted U-pattern is evident just when we consider ``small'' bin sizes.

In the case of relative prices, the volatilities also  
exhibit the same kind of intra-day pattern (figure~\ref{fig:Fig9}), but contrary with the returns, it is independent of the size of the bin, and the index we consider, but characteristic for each index. We suggested in section $6$ how this bin size independence of intra-day patterns in relative prices could be used in order to characterize ``atypical days'' for indexes and ``anomalous behaviours'' in stocks. This was showed in figures~\ref{fig:Fig13} and~\ref{fig:Fig14} where we presented our intra-day seasonalities for the mean (a) and volatility (b) in blue and the respective the cross-sectional moments for 3 days (and the single stock moments for 3 stocks) randomly picked in clear blue and we saw how the average behaviour of their moments move along with our intra-day patterns which was not the case for the day $11$ and the stock $228$.

\section*{Acknowledgements}
Esteban Guevara thanks Anirban Chakraborti, Frederic Abergel, Remy Chicheportiche and Khashayar Pakdaman for their support and discussions. Special thanks to the European Commission, the Ecuadorian Government and the Secretar\'ia Nacional de Educaci\'on Superior, Ciencia, Tecnolog\'ia e Innovaci\'on, SENESCYT.


\end{document}